\documentclass[acmsmall]{acmart}

\usepackage{graphicx}
\usepackage{amsmath}
\usepackage{amsthm}
\usepackage{xcolor}
\usepackage{hyperref}
\usepackage{booktabs}
\usepackage{caption}
\usepackage{subcaption}
\usepackage{tabularx}
\usepackage[mathscr]{euscript}
\usepackage{xspace}
\usepackage[ruled,commentsnumbered,linesnumbered,noend]{algorithm2e}
\usepackage{tikz}
\usepackage[inline]{enumitem}
\usepackage[super]{nth}
\usepackage{pgfplots}
\usepackage{tcolorbox}
\usepackage{thm-restate}
\usepackage{listings}
\usepackage{wrapfig}
\usepackage{dashrule}

\usepackage{lipsum}
\let\euscr\mathscr
\graphicspath{ {images/} }

\pgfplotsset{width=10cm,compat=1.9}
\usepgfplotslibrary{statistics}

\theoremstyle{definition}
\newtheorem{theorem}{Theorem}[section]
\newtheorem{definition}[theorem]{Definition}

\newtheorem{lemma}[theorem]{Lemma}
\newtheorem{proposition}[theorem]{Proposition}

\newtheorem{problem}{Problem}
\newtheorem*{problem*}{Problem}
\newtheorem*{remark*}{Remark}

\newenvironment{myparagraph}[1]{\vspace{0.1in}\noindent{\bf #1.}}{}

\definecolor{specBoxOutlineColor}{rgb}{0.122, 0.435, 0.698}% Rule colour
\newcommand{\specBox}[1]{%
\begin{tcolorbox}[colframe=specBoxOutlineColor,boxrule=0.5pt,arc=4pt,
      left=6pt,right=6pt,top=2pt,bottom=2pt,boxsep=0pt,width=\columnwidth,fontupper=\footnotesize]%
      {\emph{#1}}
\end{tcolorbox}%
}

\newcommand{\figlabel}[1]{\label{fig:#1}}
\newcommand{\figref}[1]{Fig.~\ref{fig:#1}}
\newcommand{\seclabel}[1]{\label{sec:#1}}
\newcommand{\secref}[1]{Section~\ref{sec:#1}}

\newcommand{\deflabel}[1]{\label{def:#1}}
\newcommand{\defref}[1]{Definition~\ref{def:#1}}

\newcommand{\thmlabel}[1]{\label{thm:#1}}
\newcommand{\thmref}[1]{Theorem~\ref{thm:#1}}
\newcommand{\proplabel}[1]{\label{prop:#1}}
\newcommand{\propref}[1]{Proposition~\ref{prop:#1}}
\newcommand{\lemlabel}[1]{\label{lem:#1}}
\newcommand{\lemref}[1]{Lemma~\ref{lem:#1}}
\newcommand{\corlabel}[1]{\label{cor:#1}}
\newcommand{\corref}[1]{Corollary~\ref{cor:#1}}

\newcommand{\tablabel}[1]{\label{tab:#1}}
\newcommand{\tabref}[1]{Table~\ref{tab:#1}}
\newcommand{\applabel}[1]{\label{app:#1}}
\newcommand{\appref}[1]{Appendix~\ref{app:#1}}
\newcommand{\algolabel}[1]{\label{algo:#1}}
\newcommand{\algoref}[1]{Algorithm~\ref{algo:#1}}

\newcommand{\set}[1]{\{#1\}}
\newcommand{\setpred}[2]{\set{#1 \,|\, #2}}
\newcommand{\tuple}[1]{\langle#1\rangle}
\newcommand{\proj}[2]{#1|_{#2}}
\renewcommand{\emptyset}{\varnothing}

\newcommand{\nats}{\mathbb{N}}
\newcommand{\rats}{\mathbb{Q}}
\newcommand{\true}{{\sf true}}
\newcommand{\false}{{\sf false}}

\newcommand{\entails}{\vdash}

\newcommand{\procs}{\euscr{P}}
\newcommand{\vals}{\euscr{V}}
\newcommand{\methods}{\euscr{M}}
\newcommand{\Times}{\euscr{T}}
\newcommand{\hist}{H}
\newcommand{\lin}{\ell}
\newcommand{\empmethods}{{\methods}^{\mathsf{\empval}}}
\newcommand{\addmethods}{{\methods}^{\mathsf{Add}}}
\newcommand{\removemethods}{{\methods}^{\mathsf{Remove}}}

\newcommand{\methodAttr}{{\sf m}\xspace}
\newcommand{\procAttr}{{\sf p}\xspace}
\newcommand{\valAttr}{{\sf v}\xspace}

\newcommand{\invTimeAttr}{{\sf inv}\xspace}
\newcommand{\resTimeAttr}{{\sf res}\xspace}

\newcommand{\methodOf}[1]{\methodAttr(#1)}
\newcommand{\procOf}[1]{\procAttr(#1)}
\newcommand{\valOf}[1]{\valAttr(#1)}

\newcommand{\invTimeOf}[1]{\invTimeAttr(#1)}
\newcommand{\resTimeOf}[1]{\resTimeAttr(#1)}
\newcommand{\opr}[2]{{#1}(#2)}
\newcommand{\abs}[1]{{\sf abs}(#1)}

\definecolor{ADTSetColor}{rgb}{0.0, 0.525, 0.043}
\newcommand{\ADT}{\mathcal{D}}
\newcommand{\registerDS}{\texttt{\textcolor{ADTSetColor}{register}}\xspace}
\newcommand{\mutexDS}{\texttt{\textcolor{ADTSetColor}{mutex}}\xspace}
\newcommand{\setDS}{\texttt{\textcolor{ADTSetColor}{set}}\xspace}
\newcommand{\stackDS}{\texttt{\textcolor{ADTSetColor}{stack}}\xspace}
\newcommand{\queueDS}{\texttt{\textcolor{ADTSetColor}{queue}}\xspace}
\newcommand{\pqueueDS}{\texttt{\textcolor{ADTSetColor}{priority-queue}}\xspace}
\newcommand{\dequeueDS}{\texttt{\textcolor{ADTSetColor}{dequeue}}\xspace}
\newcommand{\ADTSet}{\set{\setDS, \stackDS,\queueDS,\pqueueDS}}
\newcommand{\ADTReducedSet}{\set{\stackDS,\queueDS,\pqueueDS}}

\newcommand{\spec}{\mathbb{T}}
\newcommand{\RegisterSpec}{{\spec}_{\registerDS}}

\newcommand{\SetSpec}{{\spec}_{\setDS}}
\newcommand{\StackSpec}{{\spec}_{\stackDS}}
\newcommand{\QueueSpec}{{\spec}_{\queueDS}}
\newcommand{\PQueueSpec}{{\spec}_{\pqueueDS}}

\definecolor{methodNameColor}{rgb}{0.122, 0.435, 0.698}
\newcommand{\methodName}[1]{{\tt \textcolor{methodNameColor}{#1}}\xspace}
\definecolor{valueColor}{rgb}{0.698, 0.111, 0.111}
\newcommand{\val}[1]{{\tt\small \textcolor{valueColor}{#1}}\xspace}
\newcommand{\empval}{\circledcirc}
\newcommand{\mempty}{\methodName{empty}}
\newcommand{\methodsRegisters}{\methods_{\registerDS}}
\newcommand{\mread}{\methodName{read}}
\newcommand{\mwrite}{\methodName{write}}
\newcommand{\methodsQueue}{\methods_{\queueDS}}
\newcommand{\methodsPQueue}{\methods_{\pqueueDS}}
\newcommand{\enq}{\methodName{enq}}
\newcommand{\deq}{\methodName{deq}}
\newcommand{\peek}{\methodName{peek}}
\newcommand{\emp}{\methodName{empty}}
\newcommand{\methodsStack}{\methods_{\stackDS}}
\newcommand{\push}{\methodName{push}}
\newcommand{\pop}{\methodName{pop}}
\newcommand{\pass}{ok}
\newcommand{\fail}{fail}
\newcommand{\methodsSet}{\methods_{\setDS}}
\newcommand{\minsert}{\methodName{insert}}
\newcommand{\minsertOk}{\minsert_{\methodName{\pass}}}
\newcommand{\minsertFail}{\minsert_{\methodName{\fail}}}

\newcommand{\contains}{\methodName{contains}}
\newcommand{\remove}{\methodName{delete}}
\newcommand{\removeOk}{\remove_{\methodName{\pass}}}
\newcommand{\removeFail}{\remove_{\methodName{\fail}}}
\newcommand{\containsTrue}{\contains_{\methodName{\true}}}
\newcommand{\containsFalse}{\contains_{\methodName{\false}}}

\newcommand{\drop}[1]{{\sf drop}(#1)}
\newcommand{\uses}{{\sf uses}}
\newcommand{\touches}{{\sf touches}}
\newcommand{\opSort}{\mathsf{Operations}}
\newcommand{\methodSort}{\mathsf{Methods}}
\newcommand{\valSort}{\mathsf{Values}}
\newcommand{\befrel}{\mathtt{before}}
\newcommand{\bef}[1]{\befrel(#1)}
\newcommand{\mthd}{\mathtt{method}}
\newcommand{\vl}{\mathtt{value}}
\newcommand{\mdl}{\mathcal{A}}
\newcommand{\interp}{\mathcal{I}}

\newcommand{\getperm}{{\tt getPermissive}}

\newcommand{\rmsub}{{\tt removeSubhistory}}
\newcommand{\init}{{\tt initialize}}

\newcommand{\queryMin}{{\tt queryMin}}
\newcommand{\updateRange}{{\tt updateRange}}
\newcommand{\optree}{{\tt OprSet}}
\newcommand{\specialsegtree}{{\tt partitionSet}}
\newcommand{\segtree}{{\tt intervalSet}}
\newcommand{\potbotval}{{\tt PotBotVals}}
\newcommand{\potfrontval}{{\tt PotFrontVals}}
\newcommand{\potlowval}{{\tt PotLowVal}}
\newcommand{\events}{{\tt Events}}
\newcommand{\maxinvevents}{{\tt MaxInvEvents}}
\newcommand{\minresevents}{{\tt MinResEvents}}
\newcommand{\pendingreturns}{{\tt PendingReturns}}
\newcommand{\waitingreturns}{{\tt WaitingReturns}}
\newcommand{\first}{\methodName{first}}
\newcommand{\second}{\methodName{second}}

\newcommand{\propone}{\mathtt{frontEnq}}
\newcommand{\proptwo}{\mathtt{frontPeekDeq}}

\SetDataSty{textsc}

\SetKwProg{myfun}{function}{}{}
\SetKwProg{myproc}{procedure}{}{}
\SetKwProg{myhandler}{handler}{}{}
\SetKwFunction{init}{initialize}
\SetKwFunction{GetLinP}{GetLinPreservingValue}
\SetKwFunction{EmptyCheck}{EmptyCheck}
\SetKwFunction{Compliant}{InvocationResponseTimeCompliant}
\SetKwFunction{CheckLin}{CheckLin}
\SetKwFunction{CheckLinPreserving}{IsLinPreserving}
\SetKwFunction{LinOracle}{LinOracle}
\SetKwFunction{MutexLin}{CheckLin$_\mutexDS$}
\SetKwFunction{SetLin}{CheckLin$_\setDS$}
\SetKwFunction{StackLin}{CheckLin$_\stackDS$}
\SetKwFunction{DequeLin}{CheckLin$_\dequeueDS$}
\SetKwFunction{QueueLin}{CheckLin$_\queueDS$}
\SetKwFunction{PQLin}{CheckLin$_\pqueueDS$}
\SetKwFunction{MPV}{MaxPermVals}
\SetKwFunction{PBV}{GetPotBotVal}
\SetKwFunction{GPFV}{GetPotFrontVal}
\SetKwFunction{GPFVE}{NextFrontEnq}
\SetKwFunction{GPFVD}{NextFrontPeekDeq}
\SetKwFunction{RemoveOrAdd}{RemoveOrAdd}
\SetKwInput{Input}{Input}
\SetKwInOut{Output}{Output}
\SetKw{Let}{let}
\SetKw{Break}{break}
\SetKw{Continue}{continue}
\SetKw{AND}{and}
\SetKw{OR}{or}
\SetKw{NOT}{not}
\SetKw{declare}{declare}
\SetKw{report}{report}
\SetKw{exit}{exit}
\SetKwFor{Foreach}{for each}{}{}%
\SetKwFor{RepTimes}{repeat}{times}{end}
\SetKwData{NULL}{null}
\DontPrintSemicolon

\newcommand{\ucomment}[1]{\textcolor{red}{#1}}

\newcommand{\drawoper}[5][\small]{
    \draw (#2,#4) -- (#3,#4) node[midway,above] {#1 #5};
    \draw[fill=white] (#2,#4) circle (0.1);
    \draw[fill=white] (#3,#4) circle (0.1);
}
\newcommand{\drawhalfoper}[5][\normalsize]{ % Default size is \normalsize
    \draw (#2,#4) -- (#3,#4) node[midway,above] {#1 #5};
    \draw[fill=white] (#2,#4) circle (0.1);
}
\newcommand{\drawtime}[5][\normalsize]{ % Default size is \normalsize
    \draw[dashed] (#4,#2) -- (#4,#3) node[below] {#1 #5};
}
\newcommand{\drawtimeline}[3]{
    \draw[dashed] (#3,#1) -- (#3,#2);
}
\newsavebox{\figholder}
\newcommand{\wrapbox}[3]{
\sbox{\figholder}{#2}
\begin{wrapfigure}{#1}{\wd\figholder}
\usebox{\figholder}
\vspace{-0.1in}
#3
\end{wrapfigure}
}

\newcommand{\timeout}{\textsf{TO}}
\newcommand{\fastlin}{\texttt{LinP}\xspace}

\newcommand{\inferrule}[2]{\frac{\begin{array}{c}#1\end{array}}{\begin{array}{c}#2\end{array}}}
\newcommand{\inferrulena}[3]{\inferrule{#1}{#2}\;\normalfont\scriptsize\textsc{(#3)}}

\newcommand{\lts}{\mathsf{LTS}}
\newcommand{\ltstrans}[1]{\xrightarrow{#1}}
\newcommand{\ltslabs}{\Lambda}

\newcommand{\np}{\textsf{NP}}

\newcommand*\circledsmall[1]{\tikz[baseline=(char.base)]{
  \node[shape=circle,draw=black,fill=orange!20!white,inner sep=0.5pt, solid] (char) {\textcolor{black}{\tt\footnotesize{#1}}};}}

\setcopyright{cc}
\setcctype{by}
\acmDOI{10.1145/3763123}
\acmYear{2025}
\acmJournal{PACMPL}
\acmVolume{9}
\acmNumber{OOPSLA2}
\acmArticle{345}
\acmMonth{10}
\received{2025-03-25}
\received[accepted]{2025-08-12}

\begin{document}

\title{Efficient Decrease-and-Conquer Linearizability Monitoring}
\subtitle{Log-Linear time Algorithms for Unambiguous Set, Queue, Stack and Priority Queue Histories}

\author{Zheng Han Lee}
\email{zhlee@u.nus.edu}
\orcid{0009-0000-7130-2493}
\affiliation{%
  \institution{National University of Singapore}
  \city{Singapore}
  \country{Singapore}
}

\author{Umang Mathur}
\email{umathur@comp.nus.edu.sg}
\orcid{0000-0002-7610-0660}
\affiliation{%
  \institution{National University of Singapore}
  \city{Singapore}
  \country{Singapore}
}



%!TEX root=./main.tex

\begin{abstract}
Linearizability has become the de facto standard for specifying correctness 
of implementations of concurrent data structures.
While formally verifying such implementations remains challenging,
\emph{linearizability monitoring} has emerged as a promising first step to 
rule out early problems in the development of custom implementations,
and serves as a key component in approaches that stress test such implementations.
In this work, we undertake an algorithmic
investigation of the linearizability monitoring problem, which asks to check
if an execution history obtained from a concurrent data structure implementation
is linearizable.

While this problem is largely understood to be intractable in general,
a systematic understanding of when it becomes tractable has remained elusive.
We revisit this problem and first present a unified `decrease-and-conquer'
algorithmic framework for designing linearizability monitoring.
At its heart, this framework asks to identify
special \emph{linearizability-preserving} values in a given history --- 
values whose presence yields an equi-linearizable sub-history 
(obtained by removing operations of such values), 
and whose absence indicates non-linearizability.
More importantly, we prove that a polynomial time algorithm for the 
problem of identifying linearizability-preserving values,
immediately yields a polynomial time algorithm for the linearizability
monitoring problem, while conversely, intractability 
of this problem implies intractability of monitoring. 

We demonstrate the effectiveness of our decrease-and-conquer framework by instantiating it for 
several popular concurrent data types --- 
registers, sets, stacks, queues and priority queues ---
deriving polynomial time algorithms for them,
under the (\emph{unambiguity}) restriction that 
each insertion to the underlying data structure
adds a distinct value.
We further optimize these algorithms to achieve 
log-linear running time through the use of  efficient data structures
for amortizing the cost of solving induced sub-problems. 
Our implementation and evaluation on publicly available implementations of concurrent 
data structures show that our approach scales to very 
large histories and significantly outperforms existing state-of-the-art tools.
\end{abstract}


\maketitle

%!TEX root=./main.tex

\section{Introduction}
\seclabel{intro}

Linearizability, originally proposed by Herlihy and Wing~\cite{Wing1990},
serves as a systematic correctness criterion for implementations of
concurrent data structures, and has become the most popular choice of specifications.
Conceptually, it asks --- given an implementation, are all its behaviors 
equivalent to an ideal, but sequential implementation?
While full-fledged formal verification of concurrent data structure implementations 
is desirable, as with other interesting properties about programs, 
it is an undecidable property in its full generality~\cite{Bouajjani2013},
and decidable classes of programs and sequential specifications remain 
limited~\cite{Alur2000,Hamza2015,bouajjani2018reducing,vcerny2010model}.

In this work, we investigate the \emph{linearizability monitoring} problem,
a pragmatic take on the correctness problem for data structure implementations.
Instead of ensuring that all behaviors are correct,
the linearizability monitoring problem has a more modest ask --- given an execution history
obtained from the  implementation of a concurrent data structure, is this history
equivalent to some execution obtained from a sequential implementation?
Linearizability monitoring forms a fundamental component of stress testing approaches
and efficient algorithms for solving this directly translate to efficiency in the testing process.
This is indeed the theme of this work --- how efficiently
can a given history be monitored against linearizability?

As such, the complexity of checking linearizability of a history
crucially relies on the complexity of the sequential specification itself, but
can be solved in $\np$ with access to an oracle that checks for membership in the sequential specification.
Gibbons and Korach~\cite{Gibbons1997} undertook a systematic
complexity-theoretic investigation of linearizability monitoring against
the register data structure --- objects are registers and expose their $\mwrite$ and $\mread$
APIs. They showed that, in general, the monitoring 
problem is $\np$-complete for registers, and further proposed restrictions on register histories
for which the linearizability monitoring problem can be solved in polynomial time.
Since then, some works have studied popular data structures like 
stacks and queues~\cite{bouajjani2018reducing,Emmi2015,Gibbons2002LinJournal,Gibbons1999Lin}
but a unified tractability result has remained a hard challenge.
Recently though, Emmi and Enea~\cite{Emmi2018}
 proposed a wide class of data structures --- collection types --- which includes registers, queues, sets and stacks, 
and seemingly settled the tractability question for such data structures ---
they show that under the well-studied \emph{unambiguity} restriction~\cite{Enea2017,Emmi2015,bouajjani2018reducing,Gibbons1997,Emmi2018}
(each value is inserted at most once in the data structure), monitoring can be solved
in polynomial time, by first reducing it, in polynomial time, to the satisfiability
problem of a predicate logic formula, 
which can then be transformed, in polynomial time, to a Horn-SAT instance 
using a technique they call \emph{Hornification}.

We first re-investigate this result and 
identify errors in it, owing to the translation to Horn-SAT,
and show that, as such, there may not be an easy fix; see~\secref{collection-types} for more details.
As a result, the tractability problem for collection types continues to be open,
even for the unambiguity restriction.
As our main contribution in this work, \circledsmall{C1} \emph{we show that
for the class of data structures $\ADTSet$,
linearizability of unambiguous histories can, 
in fact, be monitored in log-linear time!}
To the best of our knowledge, our work proposes the most
efficient algorithms for this class of data structures.
Our work bears similarity with the work of 
Gibbons et al~\cite{Gibbons2002LinJournal,Gibbons1999Lin} 
as well as concurrent work by Abdulla et al~\cite{Abdulla2025},
but differs from them significantly, as we discuss in \secref{closely-related}.

Our algorithms are based on interesting insights.
Taking the \stackDS data structure as an example, our algorithm, at a high level,
works in a greedy fashion.
At every step, it attempts to identify 
a \emph{value} $v$ that could potentially appear at the bottom of the underlying stack
in some equivalent sequential stack history (if one exists), 
and removes all operations associated with this value $v$.
In doing so, it ensures the key invariant of \emph{linearizability preservation} --- if the
history was linearizable, then removing the operations associated with $v$ 
results in a linearizable sub-history,
and, conversely, if the resulting sub-history is linearizable, then
the original one was too.
Interestingly, our algorithms for \setDS, \queueDS and \pqueueDS also work
within this same template --- identify a linearizability-preserving value, 
remove all operations of that value and recurse.
Drawing inspiration from these algorithms, as part of our
second main contribution, \circledsmall{C2} \emph{we propose 
a generic value-centric decrease-and-conquer algorithm for linearizability monitoring,
and show that this algorithm is sound and complete for the class of container types}.
Container types are data structures whose sequential specifications are 
downward-closed with respect to the removal of all operations of any single value.
While this decrease-and-conquer algorithmic template does not, in itself,
lend to tractable monitoring algorithms, it provides a clean principle
for the design of new monitoring algorithms.
We further substantiate its utility
by instantiating it to derive our polynomial time algorithms for 
checking linearizability of unambiguous
\setDS, \stackDS, \queueDS and \pqueueDS histories, 
as well as for unambiguous \registerDS histories~\cite{Gibbons1997}.
These algorithms form precursors to our final log-linear time algorithms,
which can be obtained by further optimizations and amortizations using
classic data structures.

For our third main contribution, 
\circledsmall{C3}
\emph{we implemented our algorithms in our tool \fastlin (available publicly at~\cite{fastlintool}) and show that it
performs significantly faster than existing linearizability monitoring tools} and scales
to a million operations under a second, leveraging the  efficiency of our algorithms.

% \ucomment{Organization. Also talk about  common tool, setup etc.}

%!TEX root=./main.tex

\section{Preliminaries}
\seclabel{prelim}

In this section, we present background necessary for our presentation
including concurrent histories and operations,
as well as the linearizability monitoring problem.

\subsection{Histories and Operations}

\myparagraph{Operations}
The focus of this work is to design algorithms for checking 
linearizability of concurrent histories. 
A history essentially records concurrent operations on
data structures implementing abstract data types (ADTs).
We can model each \emph{operation} as a tuple 
$o = \tuple{id, p, m, v, t_{\invTimeAttr}, t_{\resTimeAttr}}$. 
Here, $id$ is a unique identifier for the operation $o$, 
$p \in \procs$ denotes the process executing $o$,
$m \in \methods$ denotes the method of the underlying ADT,
the component $v$ is either a value $v \in \vals$
denoting the argument of the operation $o$
or is the dedicated symbol $\empval \not\in \vals$ (in the case of methods like $m = \mempty$)
% $ret \in \vals$ denotes the return value of the operation $o$,
 and $t_{\invTimeAttr}, t_{\resTimeAttr} \in \rats_{\geq 0}$ 
denote the (rational) time corresponding to the invocation and 
response of the operation $o$;
we require that $t_{\invTimeAttr} < t_{\resTimeAttr}$. 
We will use $\procOf{o}$, $\methodOf{o}$, $\valOf{o}$, 
% $\returnOf{o}$, 
$\invTimeOf{o}$ and $\resTimeOf{o}$ 
to denote respectively the process $p$, method $m$, argument $v$,
 % return value $ret$, 
invocation time $t_{\invTimeAttr}$ and response time $t_{\resTimeAttr}$ of operation $o$. 
We will often drop the unique identifier $id$ since it will be clear from context.
Further, we will often use a more focused shorthand
$\opr{m}{arg}$
when only the method and argument attributes of the operation in question are important. 
For example, in the context of a register object, the operation $\opr{\mread}{\val{42}}$
denotes that the value $\val{42}$ has been read (or \emph{loaded}).
Likewise, for the set object, 
the operation $\opr{\minsertOk}{\val{1}}$ denotes a \emph{fresh} insertion
of the value $\val{1}$ in the set object in consideration when the set does
not already contain this value.
% 
% In the description of operations, it is common to group operations into processes for which the interval of the operations within a process, denoted by their invocation and response times, must be disjoint. In our study, we do not require the process of an operation to be identified. In fact, we may assume a system of unbounded number of processes, for which each operation may be executed independently on a separately allocated process.
%
% All operations discussed in our work have at most one value associated with an element. In the previous set example, $\minsert(1)\Rightarrow \true$ has the value $1$ associated to an element inserted into the set, while $\true$ is a return value solely to denote the successful execution of the operation. On the other hand, a stack operation $\pop()\Rightarrow 2$ has the value $2$ associated with the element being removed from the stack. We will use $\valOf{o} \in \set{\valOf{o}, \returnOf{o}}$ to denote the value associated with an element.
% \ucomment{I dont uderstand what is this.}
% 
Linearizability is a \emph{local} property~\cite{Wing1990}: 
the monitoring problem for histories with multiple objects (i.e., ADT instances) is linear-time reducible 
to the same problem for individual objects. 
We, therefore, omit object identifiers in our formalism.
Finally, we reserve the special symbol $\empval$ 
to be used in an operation $o$ which denote that the underlying data structure
is `empty' when performing $o$.
For example, in the case of \queueDS, an operation $o$ with 
$\valOf{o} = \empval$ (and $\methodOf{o} = \emp$) denotes the case when
when a client of
issues an emptiness query, or alternatively,
issues a dequeue request which fails because the underlying queue is empty.

\myparagraph{Histories}
A \emph{concurrent history} (or simply history) $\hist$ is a finite set of operations. 
For instance, the following history of a queue object
\[\hist_{\sf queue} = \set{o_1 = \tuple{p_1, \enq, \val{3}, 1, 3}, o_2 = \tuple{p_2, \deq, \val{3}, 2, 4}},\] 
comprises of two operations $o_1$ and $o_2$ performed respectively
by processes $p_1$ and $p_2$.
$o_1$ enqueues value $\val{3}$ in the invocation/response interval $[1,3]$.
$o_2$ is a dequeue operation of value $\val{3}$ by process $p_2$ in the interval $[2,4]$. 
The size $|\hist|$ of a history $\hist$ denotes the number of operations in $\hist$.
We use $\Times_\hist = \bigcup_{o\in \hist} \set{\invTimeOf{o}, \resTimeOf{o}}$ 
to denote the set of invocation and response times in $\hist$, 
and $\vals_\hist = \setpred{\valOf{o}}{o\in \hist}$ of values of operations in $H$.
We also use $\proj{\hist}{v} = \setpred{o\in \hist}{\valOf{o} = v}$ to denote the 
subhistory of $\hist$ containing exactly those operations whose value is $v$.
Likewise, $\proj{\hist}{p}$ (resp. $\proj{\hist}{m}$) denotes the sub-history
of operations in $\hist$ performed by process $p$ (resp. whose method is $m$).
A history $\hist$ is said to be \emph{sequential} if all time intervals in it are non-overlapping,
i.e., for every pair $o_1 \neq o_2 \in \hist$, we have
either $\resTimeOf{o_1} < \invTimeOf{o_2}$ or
$\resTimeOf{o_2} < \invTimeOf{o_1}$.
Histories are assumed to be well-formed, i.,e, for each process $p \in \procs$, 
the sub-history $\proj{\hist}{p}$ is sequential.

\subsection{The Linearizability Monitoring Problem}

\myparagraph{Linearizations}
A \textit{linearization} of history $\hist$ is an injective mapping 
$\lin: \hist \to \rats_{\geq 0}$ of operations in $\hist$ to timestamps such 
that for every $o \in \hist$, $\invTimeOf{o} < \lin(o) < \resTimeOf{o}$. 
In other words, a linearization is a total order of the operations 
of a history that also respects their invocation and response times. 
For the queue history $\hist_{\sf queue}$ above, 
two of the many possible linearizations include $\lin_1 = \set{o_1 \mapsto 2.5, o_2 \mapsto 3.5}$ and
$\lin_2 = \set{o_1 \mapsto 2.75, o_2 \mapsto 2.5}$.
Of course, $\lin_2$ does not meet the \emph{sequential specification} of a queue object;
see next.

\myparagraph{Sequential Specifications}
The sequential specification of an abstract data type (ADT) $\ADT$, with methods $\methods_{\ADT}$, 
expresses all possible behaviors of an object of the ADT when the object is invoked by a sequential client. 
Formally, an \emph{abstract operation} is a pair $o = \tuple{m, v}$, where 
$m \in \methods_{\ADT}, v\in\vals$
(alternatively denoted $o = \opr{m}{v}$).
% For an operation $o$,
% we use $\abs{o}$ to denote the abstract operation $\tuple{\methodOf{o}, \valueOf{o}}$.
An \emph{abstract sequential history} is a finite sequence
$\tau = o_1 \cdot o_2 \cdots o_n$ of abstract operations.
The \emph{sequential specification} 
$\spec_{\ADT}$ of an abstract data type 
$\ADT$ is a prefix-closed set of abstract sequential histories
and captures the intuitive operational meaning of the ADT.
As an example, the sequential specification $\QueueSpec$ of the queue ADT
will include the sequence 
$\tau_1 = \opr{\enq}{\val{1}} \cdot \opr{\enq}{\val{2}} \cdot \opr{\deq}{\val{1}} \cdot \opr{\deq}{\val{2}}$
but exclude the sequence 
$\tau_2 = \opr{\enq}{\val{1}} \cdot \opr{\deq}{\val{2}} \cdot \opr{\enq}{\val{2}}$. 
We will present precise sequential specifications for the concrete
data types considered in the paper in subsequent sections.
% We will often use the notation $\methods_{\ADT}$ to denote the subset of methods
% used by the data structure $\ADT$ in consideration.

\myparagraph{Linearizability}
A linearization $\lin$ of a history $\hist$ naturally induces an abstract sequential 
history, namely the sequence $\tau_\lin = \abs{o_1} \cdot \abs{o_2} \cdots \abs{o_n}$ 
corresponding to the unique total order
on operations of $\hist$ such that $\lin(o_i) < \lin(o_{i+1})$ for every $1 \leq i < n$.
Here, $\abs{o}$ denotes the abstract operation $\tuple{\methodOf{o}, \valOf{o}}$
corresponding to the operation $o$.
We say that linearization $\lin$ of history $\hist$ is a \emph{legal linearization}
with respect to an ADT specification $\spec_{\ADT}$
if the abstract sequential history $\tau_\lin$ (as defined above) 
satisfies $\tau_\lin \in \spec_{\ADT}$.
If $\lin$ is a legal linearization, then we also abuse notation
and write $\lin \in \spec_{\ADT}$.
% Let $o_1, o_2, ..., o_n$ be the sequence of operations ordered by linearization $l$ of a history. 
% Then the sequence $\tuple{\methodOf{o_1}, \valOf{o_1}}, \tuple{\methodOf{o_2}, \valOf{o_2}}, ..., \tuple{\methodOf{o_n}, \valOf{o_n}}$ of abstract operations is said to be the \emph{derived sequence} of $l$. A linearization $l$ of a history $H$ is said to be \emph{legal} with respect to a sequential specification $T$ when the derived sequence of $l$ is in $T$.

\begin{comment}
\ucomment{Remove?}
This formal definition of sequential specification, although precise, does not dictate the notation used to define such a set. In this paper, we identify a partial order on values of a sequence of abstract operations. Then we characterize the membership of the sequence in the sequential specification based on the partial order. Intuitively, the partial order characterizes the spatial and temporal order of any pair of values in the ADT. For the case of priority queues, where spatial order does not make much sense, we relate the temporal constraints with the priority constraints via the partial order.
\end{comment}

We now define the \emph{linearizability monitoring} problem, also referred to as
the problem of \emph{verifying linearizability} in ~\cite{Gibbons1997}.

\begin{definition}[Linearizable History]
Let $\spec_{\ADT}$ be a sequential specification 
corresponding to some abstract data type $\ADT$.
A concurrent history $\hist$ (derived from, say, an actual concurrent implementation of $\ADT$),
is said to be \emph{linearizable} with respect to $\spec_{\ADT}$
if there is a linearization $\lin$ of $\hist$ such that $\lin \in \spec_{\ADT}$.
\end{definition}

\begin{problem}[Monitoring Linearizability]
Let $\spec_{\ADT}$ be a sequential specification 
corresponding to some abstract data type $\ADT$.
The linearizability monitoring problem against $\spec_{\ADT}$
asks if a given input concurrent history $\hist$ is linearizable.
\end{problem}

\subsection{LTSs for sequential specifications}

% \myparagraph{LTSs for sequential specifications}
While the sequential specifications for many ADTs are often
well understood, it will be helpful for the rigor of our presentation
to also formulate them precisely.
For many popular ADTs, including all of the ones we consider
in our work, a natural form of specification is through the use of
a \emph{labeled transition system} (LTS)
where the states of the system essentially capture the internal state
of the ADT, labels on transitions represent valid actions,
and valid paths in the LTSs directly translate to abstract sequential histories.
Formally, for an ADT $\ADT$ over methods $\methods$ and values $\vals$, an LTS for its sequential specification,
has the form ${\sf LTS}_\ADT = (S_\ADT, s_0, \rightarrow)$, where $S_\ADT$ is a set of states,
intuitively representing the set of internal states of a typical sequential implementation of $\ADT$;
$s_0 \in S$ is the initial state.
The transition relation has the form $\rightarrow \subseteq S_\ADT \times \ltslabs_{\ADT} \times S_\ADT$,
where transitions are labeled from the set $\ltslabs_{\ADT} = \setpred{m(v)}{m\in \methods, v \in \vals}$
of abstract operations of $\ADT$;
we will also write $(s \ltstrans{m(v)} s')$ to denote a transition triple
$(s, m(v), s')\in \rightarrow$.
A run of such an LTS is a finite sequence 
$\rho = (q_0 \ltstrans{m_0(v_0)} q'_0) \cdot (q_1 \ltstrans{m_1(v_1)} q'_1) \cdots (q_k \ltstrans{m_k(v_k)} q'_k)$ such that $q_0 = s_0$,
for each $i < k$, we have $q'_i = q_{i+1}$ and for each $i \leq k$,
$(q_i \ltstrans{m_i(v_i)} q'_i) \in \rightarrow$.
The labeling of a run $\rho$ is the abstract sequential history
denoted by the sequence $\lambda(\rho) = m_0(v_0) \cdot m_1(v_1) \cdots m_k(v_k)$ of abstract
operations in $\rho$.
In this sense, the sequential specification of the ADT $\ADT$ presented as $\lts_\ADT$
is precisely the set $\spec_\ADT = \setpred{\lambda(\rho)}{\rho \text{ is a run of } \lts_\ADT}$.
Consider, for example the sequential specification for the \registerDS ADT
below, whose methods are $\mread$ and $\mwrite$.
\begin{comment}
\specBox{
    \!\!\underline{\bf Register data structure}: \quad $\methodsRegisters = \set{\mread, \mwrite}$, $S_\registerDS = \vals \uplus \set{\bot}$, $s_0 = \bot$  \\
    \vspace{-0.1in}
    \noindent
   \begin{align*}
   \begin{array}{ccccc}
    \inferrulena{s = s' = v}{s \ltstrans{\opr{\mread}{v}} s'}{Read$_{\textsc{reg}}$}
   &
   \quad
   &
   \inferrulena{s' = v}{s \ltstrans{\opr{\mwrite}{v}} s'}{Write$_{\textsc{reg}}$}
   \end{array}
   \end{align*}
}
\end{comment}
\specBox{
    \!\!\underline{\bf Register data structure}: \quad $\methodsRegisters = \set{\mread, \mwrite}$, $S_\registerDS = \vals \uplus \set{\bot}$, $s_0 = \bot$  \\
    \vspace{-0.1in}
    \noindent
   \begin{align*}
   \begin{array}{ccccc}
    \inferrule{s = s' = v \in \vals}{s \ltstrans{\opr{\mread}{v}} s'}
   &
   \quad
   &
   \inferrule{s' = v}{s \ltstrans{\opr{\mwrite}{v}} s'}
   \end{array}
   \end{align*}
}
Each judgement rule indicates a transition of $\lts_\registerDS$.
States are values (indicating the current value of the register).
A transition labeled with $\mread(v)$ is enabled only when 
the state is not $\bot$ and reflects
the value in the transition label, 
and moreover, the state after this transition
stays the same.
Likewise a write operation changes the state to the value being written to.
Observe that a run in $\lts_\registerDS$ is essentially a sequence of read and write operations
where every read is preceded by atleast one write 
and observes the value of the latest write before it.

\subsection{Complexity of Linearizability Monitoring and the Unambiguity Restriction}
\seclabel{unambiguous}

In general, solving the linearizability monitoring problem --- given a history $\hist$,
is there a linear order of its operations 
that will meet a given sequential specification $\spec_{\ADT}$? --- may require
a search over exponentially many permutations of the operations in $\hist$.
Indeed, this is likely the best one can do ---
Gibbons and Korach~\cite{Gibbons1997} showed that
the linearizability monitoring problem is \np-hard
even for the simplest data type \registerDS{s}.

Nevertheless, for the \registerDS data type, Gibbons and Korach also identify a useful 
restriction for which the problem becomes tractable --- 
the \emph{reads-from} restriction.
Here, in addition to the register history  $\hist$, the input is a reads-from
mapping $\mathsf{rf}_{\hist}$ that disambiguates,
for each read operation, the write operation (of the same value) that it must read from,
and the ask is to determine if there is a linearization
that obeys this mapping, i.e., the latest write that precedes
a read operation $o_r$ is precisely $o_r$ is $\mathsf{rf}_{\hist}(o_r)$.
Equivalently, the input is a history where each write operation has a unique value.
Gibbons and Korach show that, under this restriction,
the linearizability monitoring problem can be solved using an $O(n \log n)$ time algorithm~\cite{Gibbons1997}.

This insight has been extended to the analogous but more general
\emph{unambiguity restriction} for histories of a larger
class of data structures, beyond registers~\cite{Enea2017,Emmi2015,bouajjani2018reducing,Gibbons1997,Emmi2018},
and is also a key focus of this work.
While a more semantic definition of this restriction is possible,
we present the more intuitive and syntactic definition of this restriction here.
To be able to formally define such a restriction for each custom data type $\ADT$, 
we need to first identify a subset of methods $\addmethods_{\ADT} \subseteq \methods_{\ADT}$ 
that intuitively `add' some value in the internal
state of the underlying data structure. 
As an example, for the \stackDS ADT, we have $\addmethods_{\stackDS} = \set{\push}$
while for the \registerDS ADT, we have $\addmethods_{\registerDS} = \set{\mwrite}$.
Likewise, we need to identify a subset of methods $\removemethods_{\ADT} \subseteq \methods_{\ADT}$ 
that `remove' a value.
Continuing our previous examples, $\removemethods_{\stackDS} = \set{\pop}$
while $\addmethods_{\registerDS} = \emptyset$ since no method in the \registerDS ADT explicitly
removes the value written to a register.
We further require that $\addmethods_{\ADT} \cap \removemethods_{\ADT} = \emptyset$.

\myparagraph{Unambiguous histories}
We can now formally define the unambiguity restriction on histories as follows.
A history of ADT $\ADT$ is said to be an \emph{unambiguous}
history if for every value $v \in \vals_{\hist} \setminus \set{\empval}$, there is a unique operation that
`add's the value $v$, and also a unique (or none at all) operation
that `remove's the value $v$, i.e.,
$|\setpred{o \in \hist}{\valOf{o} = v, \methodOf{o} \in \addmethods_{\ADT}}| = 1$
and
$|\setpred{o \in \hist}{\valOf{o} = v, \methodOf{o} \in \removemethods_{\ADT}}| \leq 1$.
In other words, in an unambiguous history, the add and the remove operation
can be unambiguously identified for each value.
As an example, the \stackDS history 
$\hist_{\sf stack,1} = \set{\tuple{p_1, \push, \val{3}, 1, 3}, 
\tuple{p_2, \pop, \val{3}, 2, 4}, \tuple{p_1, \push, \val{42}, 5, 6}, 
\tuple{p_2, \pop, \val{42}, 5, 6}, \tuple{p_3, \peek, \val{3}, 7, 8}, \\ 
\tuple{p_1, \peek, \val{3}, 1, 10},
\tuple{p_1, \mempty, \empval, 1, 10}}$
is unambiguous.
But the \stackDS history $\hist_{\sf stack,2} = \set{\tuple{p_1, \push, \allowbreak \val{3}, 1, 3},
\tuple{p_2, \pop, \val{3}, 2, 4}, \tuple{p_1, \push, \val{3}, 5, 6}, \tuple{p_2, \pop, \val{3}, 5, 6}}$
is not because the value $\val{3}$ is added twice (two $\push$ operations for the same value).
Observe that, unambiguous histories permit multiple operations on a value
if they are not from $\addmethods$ or $\removemethods$ or even operations without
values.
As an example,
the unambiguous history $\hist_{\sf stack,1}$ contains two $\peek$ operations on the value $\val{3}$.
Our work differs from prior works~\cite{Abdulla2025,Gibbons2002LinJournal}
by allowing histories to contain such operations,
which can make the design of algorithms trickier (also see \secref{closely-related}).

The unambiguity restriction, as defined, is both realistic and straightforward to implement in practical settings.  
In most implementations of concurrent data structures, each inserted value can be assigned a unique identifier.  
This identifier may be derived from a combination of thread identifiers and local counters, or generated using an atomic fetch-and-add operation on a shared memory region.  
Either approach ensures the unambiguity of concurrent histories generated.

% \ucomment{Note about empty values and proposition about critical intervals}
%!TEX root=./main.tex

\section{Linearizability Monitoring for Container Types}
\seclabel{decrease-and-conquer}

In this section, we formalize our decrease-and-conquer framework, 
and
characterize a class of data types (i.e., sequential specifications)
for which this framework yields a sound and complete algorithm
for linearizability monitoring.
Recall that the we defined the `unambiguity' restriction
(\secref{unambiguous}) by generalizing the (register-specific) `reads-from' restriction 
to arbitrary data types. 
In the same spirit, our framework can also be viewed as a generalization of the algorithm
for monitoring of register histories under the reads-from restriction.
At a very high level, the algorithm for unambiguous register
histories, due to Gibbons and Korach~\cite{Emmi2018},
\emph{clusters} together operations of the same value;
the reads-from restriction ensures that each such 
cluster contains a unique write operation $o_w$ and exactly all the read
with the value $\valOf{o_w}$ of $o_w$.
Next, for each value $v \in \vals(\hist)$ in the history $\hist$,
the algorithm computes some summary metadata for the cluster of $v$
(namely an interval $\mathsf{zone}_v = [t_{\resTimeAttr}^v, t_{\invTimeAttr}^v]$ 
between the earliest response time $t_{\resTimeAttr}^v$ and the 
latest invocation time $t_{\invTimeAttr}^v$ in $\proj{\hist}{v}$),
and finally checks the compatibility of the metadata for each
pair of clusters. %(associated with values $v, v'$).

\myparagraph{An overview of our decrease-and-conquer approach}
In this work though, we propose that this algorithm (for monitoring against $\RegisterSpec$) 
can instead be formulated
in a \emph{greedy} style --- 
(1) pick a specific value $v$ (in the case of registers, picking any value suffices),
(2) check the compatibility of $v$ by comparing its metadata with that of all other values, 
(3) upon success, remove all operations of $v$ (`decrease') from consideration and recurse (`conquer') 
on the residual history, and
(4) upon failure, declare non-linearizability.
Our framework essentially generalizes this greedy algorithm.
A key ingredient here is the choice of the value to pick at each step;
a bad choice may result in an unsound or incomplete algorithm.
In \secref{lin-preserving}, we characterize this class of 
\emph{linearizability-preserving} values, 
which when picked ensure soundness and completeness.
We show that, this greedy decrease-and-conquer can seamlessly unify
linearizability monitoring algorithms for a large class of data types,
which we call `container' data types, defined in \secref{containers}.
In \secref{lin-preserving-transfer-complexity},
we outline how our decrease-and-conquer framework does not necessarily
reduce the complexity of linearizability monitoring but instead
cleanly delegates the burden of the algorithm on finding linearizability-preserving
values, which can be solved in isolation, often in polynomial time.

\subsection{Container Types}
\seclabel{containers}

Intuitively, containers are simply data types, whose sequential specifications
are \emph{downward closed} with respect to removal of values;
we define them  formally below.
Here, the notation $\proj{\tau}{V}$ denotes the 
maximal sub-sequence that only contains abstract operations whose
values are in the set $V$.

\begin{definition}[Container]
\deflabel{container-type}
A data type $\ADT$, with sequential specification $\spec_\ADT$,
is said to be a container type if for every $\tau \in \spec_\ADT$, 
and $V \subseteq \vals \uplus \set{\empval}$, we have,
$\proj{\tau}{V} \in \spec_\ADT$.
\end{definition}

First, observe that the above definition is not limited to only unambiguous histories.
Let us consider, for example, the register data structure 
(see \secref{stack} for formal definition of 
sequential specification $\StackSpec{}$ expressed using an LTS),
and consider the abstract sequential history 
$\tau_{\stackDS{}} = \opr{\mempty}{\empval} \cdot \opr{\push}{\val{1}} \cdot \opr{\push}{\val{2}} \cdot \opr{\pop}{\val{2}} \cdot \opr{\push}{\val{3}} \cdot \opr{\pop}{\val{3}} \cdot \opr{\pop}{\val{1}} \cdot \opr{\mempty}{\empval} \cdot \opr{\push}{\val{3}} \cdot \opr{\pop}{\val{3}} \in \StackSpec$.
Observe that the sequence obtained by removing $\val{3}$ is also a stack history, i.e.,
$\proj{\tau_{\stackDS{}}}{\set{\val{1}, \val{2}, \empval}} = \opr{\mempty}{\empval} \cdot \opr{\push}{\val{1}} \cdot \opr{\push}{\val{2}} \cdot \opr{\pop}{\val{2}} \cdot \opr{\pop}{\val{1}} \cdot \opr{\mempty}{\empval} \in \StackSpec$.
Likewise, for the abstract sequential history 
$\tau_{\pqueueDS} = \opr{\enq}{\val{2}} \cdot \opr{\enq}{\val{1}} \cdot \opr{\deq}{\val{2}} \cdot\opr{\enq}{\val{3}} \cdot \opr{\deq}{\val{3}} \cdot \opr{\deq}{\val{1}} \in \PQueueSpec$
of a priority queue (see \secref{priority-queue} for complete sequential specficiation),
observe that the projection  
$\proj{\tau_{\pqueueDS}}{\set{\val{1}, \val{2}}} = \opr{\enq}{\val{2}} \cdot \opr{\enq}{\val{1}} \cdot \opr{\deq}{\val{2}} \cdot \opr{\deq}{\val{1}}$
also belongs to $\PQueueSpec$.
Indeed, most popular data structures, including all those discussed
later in this paper, are container types:

\begin{proposition}
\proplabel{reg-set-queue-stack-pqueue-containers}
Registers, sets, queues, stacks and priority queues are container types.
\end{proposition}

\begin{remark*}
Bouajjani~\cite{bouajjani2018reducing} use the phrase `closure under projection'
to denote a similar concept as in \defref{container-type}.
Next, the class of collection types~\cite{Emmi2018} is included
in the class of container types (\defref{container-type}) because of 
the \emph{locality} requirement of collections; we present a formal proof in 
\secref{app-containers}.
On the other hand, the data type priority queue is not a collection type
(but only a container type as outlined in \propref{reg-set-queue-stack-pqueue-containers}).
This is because the specification $\PQueueSpec{}$ violates the
\emph{parametricity} requirement of collections which asks that
every sequential history in the specification $\tau \in \spec$
satisfies $\sigma(\tau) \in \spec$ for every substitution $\sigma: \vals \to \vals$.
This does not hold for the priority queue ADT: $\tau  = \opr{\enq}{\val{1}} \cdot \opr{\enq}{\val{2}} \cdot \opr{\deq}{\val{2}} \in \PQueueSpec$, but
$\sigma_{\val{1} \mapsto \val{3}}(\tau) = \opr{\enq}{\val{3}} \cdot \opr{\enq}{\val{2}} \cdot \opr{\deq}{\val{2}} \not\in \PQueueSpec$, where the substitution $\sigma_{\val{1} \mapsto \val{3}}$ maps 
the value $\val{1}$ to $\val{3}$ but does not change $\val{2}$.
\end{remark*}

A simple, yet important consequence of \defref{container-type} is that the
downward-closure with respect to values naturally lifts to histories:

\begin{proposition}
\proplabel{cotainer-linearizability-per-value}
Let $\hist$ be a history $\hist$ of a container data type.
If $\hist$ is linearizable, then $\hist \setminus \proj{\hist}{v}$ is also linearizable for all $v \in \vals_{\hist}$.
\end{proposition}

\subsection{Linearizability-Preserving Values}
\seclabel{lin-preserving}

Recall the overall gist of our greedy approach --- at each step,
identify a specific value $v$ in the history $\hist$, 
remove all its operations to get $\hist' = \hist \setminus \proj{\hist}{v}$ and recurse.
For this approach to be sound, we must ensure that if $\hist$
was linearizable, then the residual $\hist'$ continues to be so,
and likewise, if $\hist$ was non-linearizable, then so is $\hist'$.
While the former follows naturally when the data structure is a container
(\propref{cotainer-linearizability-per-value}), the latter
does not immediately follow.
This is indeed the essence of \emph{linearizability-preserving values}
which we define next.

\begin{comment}
\begin{definition}[Linearizability-Preserving Values]
\deflabel{lin-preserving}
Let $\hist$ be a history of an ADT $\ADT$ and 
let $v$ be either a value in $\vals_{\hist}$ or is a special value $\bot \not\in \vals_\hist$.
We say that $v$ is said to be \emph{linearizability-preserving} if either
\begin{enumerate*}
    \item $\hist \setminus \proj{\hist}{v}$ is not linearizable, or
    \item $\hist$ is linearizable and $v \neq \bot$.
\end{enumerate*}
\end{definition}

\begin{definition}[Linearizability-Preserving Values]
\deflabel{lin-preserving}
Let $\hist$ be a history of an ADT $\ADT$.
A linearizability-preserving value of $\hist$ is a value $v \in \vals_\hist$
such that, either
\begin{enumerate*}
    \item $\hist$ is linearizable, or
    \item $\hist \setminus \proj{\hist}{v}$ is not linearizable.
\end{enumerate*}
If no such value exists, then we say that the special symbol $\bot \not\in \vals_\hist$
is a linearizability-preserving value of $\hist$.
\end{definition}
\end{comment}

\begin{definition}[Linearizability-Preserving Values]
\deflabel{lin-preserving}
Let $\hist$ be a history of an ADT $\ADT$.
A value $v \in \vals_\hist$ is a \emph{linearizability-preserving value of $\hist$} 
if at least one of the following holds:
\begin{enumerate}
    \item $\hist$ is linearizable, or
    \item $\hist \setminus \proj{\hist}{v}$ is not linearizable.
\end{enumerate}
% Third, this definition, as alluded to previously, naturally lends
% to a recursive approach to linearizability 
% monitoring where, at each step, we remove a linearizability-preserving 
% value and check the resulting smaller history, 
% guaranteeing that the linearizability status is preserved throughout the recursion.
% If no value in $\vals_\hist$ satisfies either condition, then we say that the special symbol $\bot$ is the linearizability-preserving value of $\hist$.
When $\hist$ is non-linearizable, we also deem the special symbol $\bot\not\in \vals_\hist$
to be a linearizability-preserving value.
\end{definition}

\wrapbox{r}{
  \begin{tikzpicture}[scale=0.5]
  \drawoper{1}{2}{3}{$\mwrite(1)$}
  \drawoper{5}{6}{3}{$\mread(1)$}
  \drawoper{3}{4}{2}{$\mwrite(2)$}
  \drawoper{7}{8}{2}{$\mread(2)$}
  \end{tikzpicture}
}{
\caption{Non-linearizable $\hist_2$}
\figlabel{lin-preserve-example}
}

It would be helpful to illustrate \defref{lin-preserving} using examples.
First, consider the ambiguous, effectively sequential, 
register history $\hist_1 = \set{o_1 = \tuple{p_1, \mwrite, \val{1}, 1, 2}, 
o_2 = \tuple{p_2, \mwrite, \val{2}, 3, 4}, o_3 = \tuple{p_3, \mread, \val{1}, 5, 6}, o_4 = \tuple{p_4, \mread, \val{2}, 7, 8}, o_5 = \tuple{p_5, \mwrite, \val{3}, 9, 10}, \allowbreak o_6 = \tuple{p_5, \mwrite, \val{3}, 11, 12}, o_7 = \tuple{p_5, \mread, \val{3}, 13, 14}}$
and observe that it is non-linearizable because the 
(only) linearization of the
first three operations already violate $\RegisterSpec{}$.
The value $\val{3}$ is linearizability-preserving because
the history $\hist_1 \setminus \proj{\hist_1}{\val{3}}$ remains non-linearizable.
Now consider the unambiguous non-linearizable register history
$\hist_2 = \hist_1 \setminus \proj{\hist_1}{\val{3}} = \set{o_1, o_2, o_3, o_4}$ (see \figref{lin-preserve-example}).
Observe that both the histories $\hist_2 \setminus \proj{\hist_2}{\val{1}} = \set{o_2, o_4}$,
and $\hist_2 \setminus \proj{\hist_2}{\val{2}} = \set{o_1, o_3}$ are linearizable.
In other words, neither of the values $\val{1}$ and $\val{2}$ are linearizability preserving
for $\hist_2$. 
In this case, $\bot$ is the linearizability-preserving value for $\hist_2$.

% The special symbol $\bot$ serves as a natural 
% `failure indicator' --- when no value in $\vals_\hist$ 
% can preserve non-linearizability, 
% we can conclude that $\hist$ itself must be linearizable.
% \ucomment{This previous sentence is hard to read.}
% The special symbol $\bot$ serves as a natural 
% `failure indicator' --- it is deemed linearizability-preserving
%  exactly when no value in $\vals_\hist$ 
% is  linearizability-preserve, and in this case, $\hist$ must be non-linearizable.
The special symbol $\bot$ serves as a natural 
`failure indicator' --- when $\bot$ is deemed linearizability-preserving,
 then $\hist$ must be non-linearizable.
The notion of linearizability-preserving values is designed 
to identify values whose removal is guaranteed to preserve the linearizability 
status of the history $\hist$, as we state below in \propref{sub-history-linearizability-preserving}.
When $\hist$ is linearizable, this follows from \propref{cotainer-linearizability-per-value}.
If $\hist$ is not linearizable and $v$ is a 
linearizability-preserving value, 
then $\hist \setminus \proj{\hist}{v}$ remains non-linearizable 
(by condition (2) of the definition).

\begin{proposition}
\proplabel{sub-history-linearizability-preserving}
Let $\hist$ be a history of a container type $\ADT$ and let $v (\neq \bot) \in \vals_\hist$
be a linearizability-preserving value of $\hist$.
$\hist$ is linearizable iff $\hist \setminus \proj{\hist}{v}$ is linearizable.
\end{proposition}

\subsection{From Linearizability-Preserving Values to Monitoring and Back}
\seclabel{lin-preserving-transfer-complexity}

% \begin{minipage}[t]{0.48\textwidth}
% \begin{algorithm}[H]
% \caption{\algolabel{decrease-conquer} Decrease-and-conquer linearizability monitoring for container Types}
% \myproc{\CheckLin{$\hist$}}{
% \While{$H \neq \varnothing$} {
%     $v \gets \GetLinP(\hist)$\;
%     \lIf{$v = \bot$} {\Return false}
%     $H \gets H \setminus \proj{H}{v}$
% }

% \Return true
% }
% \end{algorithm}
% \end{minipage}
% \hfill
% \begin{minipage}[t]{0.48\textwidth}
% \begin{algorithm}[H]
% \caption{\algolabel{decrease-conquer} Checking linearizability-Preserving values using a moitoring oracle}
% \myproc{\CheckLinPreserving{$\hist$, $v$}}{
% \lIf{$\LinOracle(\hist)$}{\Return true}
% \ElseIf{$\LinOracle(\hist \setminus \proj{\hist}{v})$}{\Return false}
% \lElse{\Return true}
% }
% \end{algorithm}
% \end{minipage}

\begin{algorithm}[t]
\caption{\algolabel{decrease-conquer} Decrease-and-Conquer Linearizability Monitoring for Container Types \protect\circledsmall{C2}}
\myproc{\CheckLin{$\hist$}}{
\While{$H \neq \varnothing$} {
    $v \gets \GetLinP(\hist)$\;
    \lIf{$v = \bot$} {\Return false}
    $H \gets H \setminus \proj{H}{v}$
}

\Return true
}
\end{algorithm}

\myparagraph{The decrease-and-conquer algorithm}
We now tread towards formulating our decrease-and-conquer algorithm precisely,
and summarize its key insight --- if $\hist$ is linearizable, then
one can successively find linearizability-preserving values in $\vals_\hist$;
if $\hist$ is non-linearizable, this process fails pre-maturely and $\bot \not\in \vals_\hist$ 
is then the only linearizability preserving value.
\algoref{decrease-conquer} presents our resulting decrease-and-conquer framework
and \thmref{lin-monitoring-using-lin-preserving} states its correctness, which
at this point follows immediately.
Here, the function $\GetLinP(\hist)$ returns a linearizability-preserving
value of $\hist$.

\begin{theorem}
\thmlabel{lin-monitoring-using-lin-preserving}
\algoref{decrease-conquer} returns true iff the input history $\hist$ is linearizable. \circledsmall{C2}
\end{theorem}

\myparagraph{Time complexity of linearizability monitoring}
The running time of \CheckLin is directly determined
by that of the function \GetLinP: if 
for a history with $n = |\hist|$ operations, 
$\GetLinP(\hist)$ runs in time $T(n)$, then
$\CheckLin(\hist)$ runs in time $O(n \cdot T(n))$.
Also observe that $|\vals_\hist| \in O(n)$ and thus the time of
$\GetLinP(\hist)$ is in turn only linearly more than the time for solving
the decision problem of determining if a given value $v$ is linearizability-preserving in $\hist$.
Thus, an efficient oracle for determining if $v$  is linearizability-preserving
can be used to derive an efficient linearizability algorithm.
In the following sections, we show that there exist efficient implementations of 
\GetLinP for histories of \registerDS, \setDS, \stackDS, \queueDS and \pqueueDS
data structures, under unambiguity restriction.

\myparagraph{Linearizability-preserving values for registers with unambiguity}
Let us demonstrate the above connection
by deriving the polynomial time monitorability result for unambiguous \registerDS histories~\cite{Gibbons1997}
using our notion of linearizability-preserving values.
For a value $v$ in an unambiguous register history $\hist$, let us use 
$o^v_{\mwrite}$ to be the unique write operation of $v$ 
(i.e., $\methodOf{o_\mwrite} = \mwrite$ and $\valOf{o_w} = v$), and
let $O^v_\mread = \setpred{o \in \hist}{\methodOf{o} = \mread, \valOf{o} = v}$ 
be the set of all read operations of $v$.
Let us denote $\mathsf{minRes}_v = \min\limits_{o \in \hist_v} \resTimeOf{o}$
and $\mathsf{maxInv}_v = \max\limits_{o \in \hist_v} \invTimeOf{o}$.
Following notation from~\cite{Gibbons1997}, let us denote
the set of all
\emph{forward} values to be the set
$\mathcal{F}_\hist = \setpred{v \in \vals_\hist}{\mathsf{minRes}_v < \mathsf{maxInv}_v}$
and the set of \emph{backward} values to be
$\mathcal{B}_\hist = \setpred{v \in \vals_\hist}{\mathsf{maxInv}_v < \mathsf{minRes}_v}$.
Let us use $\mathsf{Interval}_v$ to denote the interval
$[\mathsf{minRes}_v, \mathsf{maxInv}_v]$ if $v \in \mathcal{F}_\hist$,
and to denote $[\mathsf{maxInv}_{v}, \mathsf{minRes}_{v}]$ otherwise (if $v \in \mathcal{B}_\hist$).
We now characterize linearizability-preserving values for
a \registerDS history.

\begin{lemma}
\lemlabel{linearizability-preserving-registers}
Let $\hist$ be an unambiguous \registerDS history and $v \in \vals_\hist$.
If each of the following holds, then $v$ is linearizability-preserving for $\hist$:
\begin{enumerate}
    \item For every $o \in O^v_\mread$, $\invTimeOf{o^v_\mwrite} < \resTimeOf{o}$.
    \item If $v \in \mathcal{F}_\hist$, then for every $v' \neq v \in \mathcal{F}_\hist$,
    we have $\mathsf{Interval}_v \cap \mathsf{Interval}_{v'} = \emptyset$, 
    and for every $v' \in \mathcal{B}_\hist$, we have
    $\mathsf{Interval}_{v'} \not\subseteq \mathsf{Interval}_v$.
    \item If $v \in \mathcal{B}_\hist$, then for every $v' \neq v \in \mathcal{F}_\hist$,
    we have $\mathsf{Interval}_v \not\subseteq \mathsf{Interval}_{v'}$.
\end{enumerate}
If no value $v \in \vals_\hist$ meets the above conditions, then $\bot$ is linearizability-preserving for $\hist$.
\end{lemma}

It is easy to observe, based on \lemref{linearizability-preserving-registers},
that the problem --- check if a value $v$ is linearizability-preserving
in an unambiguous register history --- can be solved in time $O(n)$,
giving a straightforward $O(n^2)$ implementation of the function
\GetLinP, and thus an $O(n^3)$ algorithm for linearizability monitoring
in this setting using the template of \algoref{decrease-conquer}.
Of course, the algorithm of~\cite{Gibbons1997} is optimized further based on other 
crucial insights and runs in $O(n\log{n})$ time.

\myparagraph{Checking linearizability-preservation using a monitoring oracle}
\algoref{decrease-conquer} essentially shows that
there is a polynomial-time Turing reduction from the problem
of checking if a value is linearizability preserving to the problem of monitoring
linearizability. Is the converse also true?
We answer this in the positive.
Suppose we have access to an oracle \LinOracle that returns on history $\hist$ true iff $\hist$ is linearizable.
Then, the truth table reduction given by
the query $\LinOracle(\hist) \lor \neg \LinOracle(\hist \setminus \proj{\hist}{v})$
accurately determines if $v$ is linearizability-preserving for $\hist$.
As a consequence, both the problems of linearizability monitoring of histories as well as the problem of determining if a value is linearizability preserving are \emph{equi-tractable} for a given
data type --- either
both are solvable in polynomial time, or both are $\np$-hard.

% \ucomment{Remove this.}
% In this section, we explore the commonly used data types and their unambiguous histories to construct an efficient procedure to determine redundant values. We remark that there are likely more than one efficient approach in finding redundant values for several data types. In our case, we generally examine the properties exhibited by the operations of the values, with regards to the surrounding operations of other values, to determine whether the value must be redundant. And the absence of values exhibiting such properties necessarily imply $\bot$ to be redundant. The correctness of our approach follows naturally from the behaviour of the data types.

%!TEX root=./main.tex

\section{Standardizing Unambiguous Histories}
\seclabel{standardization}

This section is meant to be a precursor to 
\secref{set}, \secref{stack}, \secref{queue} and \secref{priority-queue}.
Here we outline some pre-processing steps that uniformly apply to input 
unambiguous histories of each data structure
in $\mathsf{ADTs} = \set{\setDS{}, \stackDS{}, \queueDS{}, \pqueueDS{}}$.
These steps can be carried out efficiently, preserve linearizability,
help us state our algorithms succinctly and sometimes also help us arrive at
overall faster algorithms.
We say that the resulting simplified histories are `standardized'.

\myparagraph{Making histories well-matched}
Recall that, the unambiguity restriction asks that the input history $\hist$
has, for each value $v \in \vals_{\hist}$, exactly one operation whose method comes from
the pre-defined set $\addmethods$ and
also at most one operation with method from the 
other pre-defined set $\removemethods$.
When for a value $v$, the history contains both such operations, we say that it is \emph{matched},
and say that $v$ is \emph{unmatched} otherwise.
We say that an unambiguous history $\hist$ is 
\underline{\emph{well-matched}} if for every $v \in \vals_\hist \setminus \set{\empval}$,
$v$ is matched in  $\hist$.

For each data structure
$\ADT \in \ADTSet$ 
we consider, there is a unique dedicated add
method $\methodAttr{}^{\mathsf{Add}}_{\ADT}$ and also a unique dedicated
remove method $\methodAttr{}^{\mathsf{Remove}}_{\ADT}$
(i.e., ${\methods}^{\mathsf{Add}}_{\ADT} = \set{\methodAttr{}^{\mathsf{Add}}_{\ADT}}$
and ${\methods}^{\mathsf{Remove}}_{\ADT} = \set{\methodAttr{}^{\mathsf{Remove}}_{\ADT}}$).
Fortunately, this means that unambiguous histories can be extended
using a simple linear time procedure so that the resulting histories are
well-matched.
For each unmatched value $v$ in an unambiguous history $\hist$ on $\ADT$,
we add an operation $o_v^{\mathsf{Remove}}$ to $\hist$ with $\valOf{o} = v$,
$\methodOf{o} = \methodAttr{}^{\mathsf{Remove}}_{\ADT}$,
$\invTimeOf{o} = t$ and $\resTimeOf{o} = t+1$,
where $t$ is some time larger than the maximum response time of any operation of $\hist$.
We observe that, this such an extension takes $O(n)$ time
(n = $|\hist|$), and for the data structures we study,
it results in a history that is linearizable iff $\hist$ is.
For the remaining of this paper, we shall assume that our histories are well-matched.

\myparagraph{Adjusting invocation and response times}
For each  $\ADT \in \ADTSet{}$,
a well-matched unambiguous history $\hist$ of $\ADT$, 
if linearizable, must linearize the unique add operation $o_v^{\mathsf{Add}}$
of a value $v$ before every other operation that touches $v$,
and likewise, must linearize the remove operation $o_v^{\mathsf{Remove}}$ 
after other operations that touch the value $v$.
With this, we say that a value $v$ is
invocation-response-time-compliant if 
for every $o\in \touches(\hist, v)$,
\begin{enumerate*}
    \item $\invTimeOf{o_v^{\mathsf{Add}}} \leq \invTimeOf{o} \leq \invTimeOf{o_v^{\mathsf{Remove}}}$, and
    \item $\resTimeOf{o_v^{\mathsf{Add}}} \leq \resTimeOf{o} \leq \resTimeOf{o_v^{\mathsf{Remove}}}$.
\end{enumerate*}
Here, the set $\touches(\hist, v)$ is the set of those operations
of $v$ that either add $v$ or use $v$ in the underlying data structure~\cite{Emmi2018}.
For the data structures $\ADT \in \set{\stackDS{}, \queueDS{}, \pqueueDS{}}$,
we have $\touches(\hist, v) = \proj{\hist}{v}$,
while for $\ADT = \setDS$, 
we have $\touches(\hist, v) = \setpred{o \in \proj{\hist}{v}}{\methodOf{o} \not\in \set{\containsFalse, \removeFail}}$, i.e., we exclude those operations $o$ that
do not actually touch $v$, even if $\valOf{o} = v$.
A well-matched unambiguous history $\hist$ is said to be
\underline{\emph{invocation-response-time-compliant}}
if every value in it is.
Observe that, when $\hist$ is invocation-response-time-compliant,
then for each $v \in \vals_\hist \setminus \set{\empval}$,
$\proj{\hist}{v}$ is linearizable, and otherwise, $\hist$ is non-linearizable.
Indeed, \algoref{compliant-algo} transforms a given input history to an equi-linearizable
history that is invocation-response-time-compliant.

%!TEX root=../main.tex

\begin{algorithm}[t]
\caption{Making a well-matched unambiguous history $\hist$ of data structure $\ADT \in \ADTSet$ invocation-response-time-compliant}
\algolabel{compliant-algo}
\myproc{\Compliant{$\hist$}}{
\For{$v \in \vals_\hist \setminus \set{\empval}$}
{
    $o_v^{\mathsf{Add}} \cdot t_{\resTimeAttr} := \min \setpred{\resTimeOf{o}}{o\in \touches(\hist, v)}$\;
    $o_v^{\mathsf{Remove}} \cdot t_{\invTimeAttr} := \max \setpred{\invTimeOf{o}}{o\in \touches(\hist, v)}$\;
    \For{$o\in \touches(\hist, v) \setminus \set{o_v^{\mathsf{Add}}, o_v^{\mathsf{Remove}}}$}
    {
        $o \cdot t_{\invTimeAttr} := \max\set{\invTimeOf{o}, \invTimeOf{o_v^{\mathsf{Add}}}}$ \;
        $o \cdot t_{\resTimeAttr} := \min\set{\resTimeOf{o}, \resTimeOf{o_v^{\mathsf{Remove}}}}$
    }
    \lIf{$\exists o \in \proj{\hist}{v}, \resTimeOf{o} \leq \invTimeOf{o}$}{\Return $\bot$}

}
\Return $\hist$
}
\end{algorithm}

\begin{lemma}
\lemlabel{invocation-response-time-compliance}
\algoref{compliant-algo} runs in $O(n)$ time on histories with $n$ operations.
Next, for $\ADT \in \ADTSet$ and for an unambiguous well-matched history $\hist$ of 
$\ADT$, if we have $\Compliant{\hist} = \bot$, then
$\hist$ is non-linearizable.
Otherwise, the history $\hat{\hist} = \Compliant{\hist}$ is invocation-response-time-compliant, and
further, $\hat{\hist}$ is linearizable iff $\hist$ is linearizable.
\end{lemma}

\noindent
\begin{proof}[\noindent \rm {\it Proof} (Sketch)]
Non-linearizability in the case $\Compliant{\hist}\allowbreak = \bot$,
invocation-response-time-compliance of $\hat{\hist}$ and running time of \algoref{compliant-algo} are straightforward to argue; we skip them here. 
We argue that $\hat{\hist}$,
and $\hist$ are equi-linearizable (in the case when $\hat{\hist} = \Compliant{\hist} \neq \bot$).
Suppose $\hist$ is linearizable, and let $\lin$ be a witnessing linearization.
Observe that if $\lin$ is not a valid linearization of $\hat{\hist}$,
then one of the following holds for some $o \in \proj{\hist}{v} \setminus \set{o_v^{\mathsf{Add}}, o_v^{\mathsf{Remove}}}$:
\begin{enumerate*}
    \item $\lin(o_v^{\mathsf{Add}}) > \resTimeOf{o} \geq \lin(o)$,
    \item $\lin(o_v^{\mathsf{Remove}}) < \invTimeOf{o} \leq \lin(o)$,
    \item $\lin(o) < \invTimeOf{o_v^{\mathsf{Add}}} \leq \lin(o_v^{\mathsf{Add}})$, or
    \item $\lin(o) > \resTimeOf{o_v^{\mathsf{Remove}}} \geq \allowbreak \lin(o_v^{\mathsf{Remove}})$.
\end{enumerate*}
Any of these, if holds, also implies that $\lin$ is not a legal linearization of $\hist$, contradicting our assumption.
Hence, $\lin$ must be a legal linearization of $\hat{\hist}$.
Now, suppose that $\hat{\hist}$ is linearizable, and let $\hat{\lin}$ be a linearization that witnesses this.
Every modification to the history in \algoref{compliant-algo} reduces an operation's interval.
Hence, $\hat{\lin}$ must also be a legal linearization of $\hist$.
The conclusion follows.
\end{proof}

\myparagraph{Removing empty operations}
Finally, we discuss how we further simplify our histories by
removing operations whose method is $\mempty$.
For every $\ADT \in \ADTReducedSet{}$,
this is the only method $m$ which satisfies ---
for each operation $o\in \hist$, $\methodOf{o} = m$ iff $\valOf{o} = \empval$.
For a history $\hist$, we will use 
$\hist_{\empval} = \setpred{o \in \hist}{\methodOf{o} = \mempty}$.
A well-matched and invocation-response-time-compliant unambiguous history $\hist$ is said to be \underline{\emph{empty-free}}
if it has no empty operations, i.e., $\hist_\empval = \varnothing$.
Observe that, given any well-matched and invocation-response-time-compliant unambiguous history $\hist$,
$\hist \setminus \hist_\empval$ is empty-free.
One may naturally ask if removing $\mempty$ operations from $\hist$ always preserves linearizability.
We now note an important property of each
of the data structure $\ADT \in \ADTReducedSet$ we consider, namely
that for every $\tau = \tau_1 \cdot o \cdot \tau_2 \in \spec_\ADT$,
if $\methodOf{o} = \mempty$, then both the sequential sub-sequences
$\tau_1, \tau_2 \in \spec_\ADT$, and further, 
$\tau_1 = \proj{\tau}{V}$ for some $V\subseteq \vals_H$.
Intuitively, we wish to avoid linearizing $\mempty$ operations
in a way that interleaves operations that touches other values (and hence indicates the presence of those values).
As it turns out, if we can do so, then we can safely remove such operations
while preserving linearizability.
The following result now characterizes when to safely remove
empty operations from input histories:

\begin{restatable}{lemma}{removeEmptyOperations}
\lemlabel{empty-remove}
Let $\ADT \in \ADTReducedSet$
Let $\hist$ be a well-matched unambiguous and invocation-response-time-compliant history of $\ADT$. 
$\hist$ and $\hist\setminus \hist_\empval$ are equi-linearizable if
for all $o \in \hist_\empval$, $[\invTimeOf{o}, \resTimeOf{o}] \nsubseteq \bigcup_{v\in \vals_\hist\setminus \set{\empval}}[\resTimeOf{o_v^{\mathsf{Add}}}, \invTimeOf{o_v^{\mathsf{Remove}}}]$.
Otherwise, $\hist$ is non-linearizable.
\end{restatable}

%!TEX root=../main.tex

\begin{algorithm}[t]
\caption{Making a well-matched and invocation-response-time-compliant unambiguous history $\hist$ of data structure $\ADT \in \ADTReducedSet$ empty-free}
\algolabel{empty-algo}
\myproc{\EmptyCheck{$\hist$}}{
${\tt BadZone} \gets \bigcup_{v\in \vals_\hist\setminus \set{\empval}}[\resTimeOf{o_v^{\mathsf{Add}}}, \invTimeOf{o_v^{\mathsf{Remove}}}]$\;
\For{$o \in \hist_\empval$}
{
    \lIf {$[\invTimeOf{o}, \resTimeOf{o}] \subseteq {\tt BadZone}$} { \Return $\false$ }
}
\Return $\true$\;
}
\end{algorithm}

\algoref{empty-algo} takes in a well-matched and invocation-response-time-compliant unambiguous history $\hist$ of $\ADT \in \ADTReducedSet$,
and returns $\true$ if removing $\mempty$ operations from $\hist$ produces an equi-linearizable history, and $\false$ otherwise.
It achieves this by performing the exact check described in \lemref{empty-remove}.

\begin{lemma}
\lemlabel{empty-remove-algo}
\algoref{empty-algo} runs in $O(n \log{n})$ time on histories with $n$ operations.
Next, for $\ADT \in \ADTReducedSet$ and for an unambiguous well-matched and invocation-response-time-compliant history $\hist$ of 
$\ADT$, if we have $\EmptyCheck{\hist} = \false$, then
$\hist$ is non-linearizable.
Otherwise, the history $\hat{\hist} = \hist \setminus \hist_\empval$ is linearizable iff $\hist$ is linearizable.
\end{lemma}

\noindent
\begin{proof}[\noindent \rm {\it Proof} (Sketch)]
We represent a possibly non-contiguous time interval as a list of disjoint intervals ${\tt BadZone} \in (\rats \times \rats)^*$,
sorted in ascending order.
With that, the union of intervals is known to be achievable in $O(n \log{n})$ time,
where $n$ is the number of intervals.
Thereafter, we can check if a given time interval $[a, b]$ is contained in the union of intervals in $L$ in $O(\log{n})$ time
by performing a binary search on ${\tt BadZone}$.
Hence, the overall running time of \algoref{empty-algo} is $O(n \log{n})$.
The correctness of \algoref{empty-algo} follows from \lemref{empty-remove}.
\end{proof}

\myparagraph{Standardized History}
Let $\ADT \in \ADTSet$.
An unambiguous history $\hist$ of $\ADT$ is said to be \underline{\emph{standardized}}
if it is well-matched, invocation-response-time-compliant and empty-free.
Based on \lemref{invocation-response-time-compliance}, \lemref{empty-remove}
and \lemref{empty-remove-algo},
we know that given an unambiguous history that is not standardized, one can
transform it into a standardized one in $O(n \log{n})$ time.
Our algorithms in Sections~\ref{sec:set}, \ref{sec:stack}, \ref{sec:queue} and \ref{sec:priority-queue}
rely on standardization as a pre-processing step.
%!TEX root=../main.tex

\begin{comment}
1. The Seq spec T_D
2. Proposition that T_D is container. Before stating the proposition, additionally remark (plain text) that it is/not a collection type
3. Subsec: Lineariz. Monitoring
    a. Explanation, with figure
    b. Lemma: Lin-preserving is polytime
    c. Corollary: Lin monitoring is polytime
    d. Optimization
        i. English description
        ii. Actual Algo
        iii. Theorem: Lin monitorng can be solved in O(nlog n) time
4. Evaluation
\end{comment}

\begin{figure}[t]
\specBox{
    \!\!\underline{\bf Set data structure}: \quad $\methodsSet = \set{\minsertOk, \minsertFail, \removeOk, \removeFail, \containsTrue, \containsFalse}$, $S_{\setDS} \subseteq_{\textsf{fin}} \vals$, $s_0 = \emptyset$ \\
    \vspace{-0.1in}
    \noindent
   \begin{align*}
   \begin{array}{ccccc}
   \inferrule{v\not\in s, s' = s \uplus \set{v}}{s \ltstrans{\opr{\minsertOk}{v}} s'}
   &
   \inferrule{v\in s, s' = s}{s \ltstrans{\opr{\minsertFail}{v}} s'}
   &
   \inferrule{v \in s, s' = s \setminus \set{v}}{s \ltstrans{\opr{\removeOk}{v}} s'}
   &
   \inferrule{v \not\in s, s' = s}{s \ltstrans{\opr{\removeFail}{v}} s'}
   \end{array}
   \end{align*}
}
\caption{Sequential specification $\SetSpec{}$ as an LTS. We skip explicitly presenting 
transitions for $\containsTrue{}$ (resp. $\containsFalse{}$) 
since they are identical to that of $\minsertFail{}$ (resp. $\removeFail{}$).
\figlabel{set-lts}}
\end{figure}

\section{The $\setDS{}$ data structure}
\seclabel{set}

The sequential specification for the set data type is presented in \figref{set-lts}.
The operation $\minsertOk$ represents addition of a fresh value to the underlying set,
whereas $\minsertFail$ represents addition of a value that is already 
present in the underlying set at the time of addition.
The operations $\removeOk$ and $\removeFail$ respectively
represent a successful (i.e., value is present in the set) and unsuccessful (i.e., value is absent) 
removal of a value.
Likewise, $\containsTrue{}$ and $\containsFalse$ are
operations corresponding to a check of containment of a value
that respectively succeed (i.e., value is present) or
fail (i.e., value is absent).
Finally the $\emp$ operation represents that the underlying set is empty, that is contains no value at all.
In the rest of this section, 
we will assume that all occurrences of $\containsTrue{}$
(resp. $\containsFalse{}$)
are replaced with $\minsertFail{}$ (resp. $\removeFail{}$),
given that they have identical semantics.
We finally note that the add and remove operations for the \setDS ADT are: 
$\methods{}_{\setDS}^{\mathsf{Add}} = \set{\minsertOk}$ and
$\methods{}_{\setDS}^{\mathsf{Remove}} = \set{\removeOk}$.

%!TEX root=../main.tex

\subsection{Linearizability-preserving values and monitoring for unambiguous $\setDS$ histories}
\seclabel{set-redundant}

For the \setDS data structure, linearizability monitoring
is very intuitive and relies on the insight that,
each value can be independently examined to determine linearizability.
Assuming histories have been standardized as in \secref{standardization},
this boils down to checking if all operations
of the form $\removeFail{}(v)$ occur outside of times when operations
we must schedule those operations that touch $v$;
\defref{safe-set} precisely formalizes this notion.
% Indeed, recall for a \setDS history $\hist$
% which is standardized (and thus invocation-response-time-compliant),
% all operations in $\touches(\hist, v) = \proj{\hist}{v} \setminus \proj{\hist}{\removeFail}$
% must have their response times earlier than the response of the unique
% $\removeOk$
\lemref{linearizability-preserving-set} states the justification of \defref{safe-set}
in terms of linearizability preservation;
here $o^v_{\minsertOk}$ and $o^v_{\removeOk}$ denote the unique operations
with value is $v$ and for whose methods are respectively, $\minsertOk$ and
$\removeOk$.

\begin{definition}[Safe values]
\deflabel{safe-set}
Let $\hist$ be an unambiguous $\setDS$ history.
A value $v \in \vals_\hist$ is said to be safe
for $\hist$ if for every $o \in \proj{\hist}{v} \cap \proj{\hist}{\removeFail}$,
we have either
$\invTimeOf{o} < \resTimeOf{o^v_{\minsertOk}}$ or $\resTimeOf{o} > \invTimeOf{o^v_{\removeOk}}$.
\end{definition}

\begin{restatable}{lemma}{SetLinearizabilityPreservingLemma}
\lemlabel{linearizability-preserving-set}
Let $\hist$ be a standardized unambiguous $\setDS$ history and let $v \in \vals_\hist$. 
If $v$ is safe for $\hist$ then it is a linearizability-preserving value for $\hist$.
If no safe value exists in $\vals_\hist$, then $\bot$ is linearizability-preserving for $\hist$.
\end{restatable}

Based on \defref{safe-set}, a simple
algorithm is evident that, in linear time, enumerate exactly which values are safe, giving us
an overall $O(n^2)$ algorithm for linearizability monitoring, based on
the template of \algoref{decrease-conquer}.
Nevertheless, an overall linear time algorithm is still
possible based on the insight that the safety status of
each value in a history remains same even when
other values are removed.
In other words, one can determine whether a history is linearizable 
by instead checking if each of its value is safe.
This gives the straightforward $O(n)$ time algorithm we show in \algoref{set-optimized}.
Finally, given that standardization can take upto $O(n \log n)$ time, we 
have the following result (we use $n$ denote the number of operations in the input history):

\begin{theorem}
  Linearizability monitoring for unambiguous \setDS histories takes $O(n \log{n})$ time. \circledsmall{C1}
\end{theorem}

%!TEX root=../main.tex

\begin{algorithm}[t]
\caption{Linearizability monitoring for unambiguous \setDS histories}
\algolabel{set-optimized}
\myproc{\SetLin{$\hist$}}{
\tcp{Let MinRes and MaxInv be hashtables with elements of $\vals_\hist$ as key}
\lForEach{$o \in \proj{\hist}{\minsertOk}$}{${\tt MinRes}[\valOf{o}] \gets \resTimeOf{o}$}
\lForEach{$o \in \proj{\hist}{\removeOk}$}{${\tt MaxInv}[\valOf{o}] \gets \invTimeOf{o}$}
\For{$o \in \proj{\hist}{\removeFail}$}
{
    \lIf{$({\tt MinRes}[\valOf{o}] < \invTimeOf{o})$ \AND $(\resTimeOf{o} < {\tt MaxInv}[\valOf{o}]$)}{\Return $\false$}
}
\Return true
}
\end{algorithm}
% \vspace{-0.1in}

In \secref{set-empirical}, we demonstrate the empirical effectiveness of our algorithm
for linearizability monitoring of unambiguous \setDS histories, and comparison
its performance with that of
prior state-of-the-art publicly available tools.

% \clearpage
%!TEX root=../main.tex

\section{The $\stackDS{}$ data structure}
\seclabel{stack}

The sequential specification for the stack data type is presented in \figref{stack-lts}
as an LTS. 
The state of the LTS essentially encodes a stack, given that
values are `\push'ed to, as well as `\peek'ed and `\pop'ped from the end of the sequence.
The empty operation $\opr{\emp}{\empval}$ can model unsuccessful 
$\push$ or $\peek$ operations, i.e., when $s = \epsilon$.
We note that the add and remove operations for the \stackDS ADT are: 
$\methods{}_{\stackDS}^{\mathsf{Add}} = \set{\push}$ and
$\methods{}_{\stackDS}^{\mathsf{Remove}} = \set{\pop}$.

\begin{figure}[t]
\specBox{
    \!\!\underline{\bf Stack data structure}: \quad $\methodsStack = \set{\push, \pop, \peek}$, $S_\stackDS = \vals^*$, $s_0 = \epsilon$ \\
    \vspace{-0.1in}
    \noindent
   \begin{align*}
   \begin{array}{ccccccc}
   \inferrule{s' = s \cdot v}{s \ltstrans{\opr{\push}{v}} s'}
   &
   \;
   &
   \inferrule{\exists s'', s = s' = s'' \cdot v}{s \ltstrans{\opr{\peek}{v}} s'}
   &
   \;
   &
   \inferrule{s = s' \cdot v}{s \ltstrans{\opr{\pop}{v}} s'}
   &
   \;
   &
   \inferrule{s = s' = \epsilon}{s \ltstrans{\opr{\emp}{\empval}} s'}
   \end{array}
   \end{align*}
}
\caption{Sequential specification $\StackSpec{}$ as an LTS.
\figlabel{stack-lts}}
\end{figure}

%!TEX root=../main.tex

\subsection{Linearizability-preserving values for unambiguous \stackDS histories}
\seclabel{stack-linearizability-preserving}

At a high level, our algorithm for monitoring $\stackDS$ histories 
adheres to our decrease-and-conquer template (\algoref{decrease-conquer}),
and at each step, identifies \emph{potentially bottom values}.
Intuitively, such a value $v$ can occur at the bottom of the stack,
in some legal linearization, if one exists. 
Consider, for example, 
\wrapbox{r}{
  \begin{tikzpicture}[scale=0.4]
  \drawoper{2}{4}{1}{$\push(\val{1})$}
  \drawoper{7}{10}{1}{$\peek(\val{1})$}
  \drawoper{15}{17}{1}{$\pop(\val{1})$}

  \drawoper{11}{13}{2}{$\push(\val{3})$}
  \drawoper{12}{14}{3}{$\pop(\val{3})$}

  \drawoper{1}{6}{2}{$\push(\val{2})$}
  \drawoper{5}{9}{3}{$\pop(\val{2})$}
  
  \drawtimeline{0.5}{5}{3}
  \drawtimeline{0.5}{5}{8}
  \drawtimeline{0.5}{5}{16}
  \end{tikzpicture}
}{\figlabel{stack-potbot}} 
the history shown on the right.   
Observe that operations on value \val{1} 
can be scheduled at the indicated linearization points, and
in doing so, operations of other values can be `shrunk' and clustered
within a single interval demarcated by these linearization points.
We treat this as an indication that the value \val{1} can 
potentially serve as the bottom-most value in a possible linearization
of the history, if one were to exist.
In contrast, value \val{2} cannot be regarded as a potentially bottom value, 
since the operation $\opr{\pop}{\val{2}}$
must be sandwiched between the operations 
$\opr{\push}{\val{2}}$ and $\opr{\pop}{\val{1}}$.
That is, the $\pop$ operation of value \val{2} is 
constrained by the guaranteed presence of value \val{1} in the stack.
In the following, we formally define such values (\defref{potentially-bottom-stack}) 
and show that they can serve as linearizability-preserving values (\lemref{linearizability-preserving-stack}).

\begin{definition}[Potentially Bottom Values]
\deflabel{potentially-bottom-stack}
Let $\hist$ be an unambiguous $\stackDS$ history. 
A value $v \in \vals_\hist$ is said to be a potentially bottom value for $\hist$
if for each $o \in \proj{\hist}{v}$,
there is a time point $t \in \rats_{\geq 0}$
with $\invTimeOf{o} < t < \resTimeOf{o}$ 
such that for all $v' \in \vals_\hist \setminus \set{v}$, we have either
\begin{enumerate*}
  \item $t < \min\setpred{\resTimeOf{o'}}{o' \in \proj{\hist}{v'}}$, or
  \item $t > \max\setpred{\invTimeOf{o'}}{o' \in \proj{\hist}{v'}}$.
\end{enumerate*}
\end{definition}

% \ucomment{revisit example and explain why that particular value in the example is indeed a potentially bottom value and why the other one is not.}

% We use $\potbotval(H)$ to denote the set of potential bottom values of the given stack history $H$. We say that the set is \emph{sound} if all values of the set are redundant, and \emph{sufficient} if the emptiness of the set implies $\bot$ to be redundant. We show that $\potbotval(H)$ is indeed sound and sufficient.

As we alluded to before, potentially bottom values are linearizability-preserving (\lemref{linearizability-preserving-stack})
This indeed aligns with the intuition behind how we defined potentially bottom values:
if $v$ is potentially bottom in $\hist$, then its operations provably
cannot interfere with the operations of other values, and removing the operations of $v$ altogether 
must not affect the admissibility of the operations of other values in the stack history.

\begin{restatable}{lemma}{stackredundant}
\lemlabel{linearizability-preserving-stack}
Let $\hist$ be a standardized unambiguous $\stackDS$ history and let $v \in \vals_\hist$. 
If $v$ is a potentially bottom value for $\hist$ then it is linearizability-preserving for $\hist$.
If no value potentially bottom for $\hist$, then $\bot$ is linearizability-preserving for $\hist$.
\end{restatable}

Let us now consider the time complexity of identifying potentially bottom values.
For a value $v$ in a standardized unambiguous \stackDS history $\hist$, 
let us denote the interval 
$(\min\setpred{\resTimeOf{o}}{o \in \proj{\hist}{v}}, \max\setpred{\invTimeOf{o}}{o \in \proj{\hist}{v}})$ 
as the \emph{critical interval of $v$}.
Based on \defref{potentially-bottom-stack}, we observe that
to determine if $v$ is a potentially bottom value,
it suffices to check that, for all $o \in \proj{\hist}{v}$, 
the interval $(\invTimeOf{o}, \resTimeOf{o})$ 
is not fully contained within the critical intervals of any other value.
Thus, an $O(n^2)$ algorithm for determining if a value $v$
is potentially bottom, is straightforward --- first compute the critical interval for each value,
and then check if the span $(\invTimeOf{o}, \resTimeOf{o})$ of
 every operation $o$ of $v$ is not entirely contained inside 
 the union of the critical intervals of other values combined.
We include a concrete implementation 
of the said procedure in \appref{stack-proof} (see \algoref{stack-redundant}).
This means linearizability can be monitored in $O(n^4)$ time.

% \algoref{stack-redundant} shows the implementation of \GetLinP for unambiguous stack histories. We return any potential bottom value, or $\bot$ if there are none. The correctness of the algorithm is guaranteed by \lemref{stack-redundant-sound} and \lemref{stack-redundant-sufficient}. It is also clear that intersection of intervals can be implemented efficiently by representing the result as a set of contiguous segments. The tractability of \algoref{stack-redundant} brings us directly the following theorem.
%!TEX root=../main.tex

%!TEX root=../main.tex

\begin{algorithm}[t]
\caption{Linearizability-preserving values for unambiguous $\stackDS$ histories}
\algolabel{stack-redundant-optimized}
\myproc{$\init(\hist)$}{
$\optree \gets \hist$, $\potbotval \gets \varnothing$ \;
$\specialsegtree.\init(\hist)$ \;
}
\BlankLine
\myproc{$\GetLinP_\stackDS(H)$}{
% \If(\tcp*[h]{one-time initialization}){$\optree = \varnothing$}{
%     $\optree \gets H$\;
%     $\specialsegtree.\init(H)$
% }
% \;
\While{$\potbotval = \varnothing$} {
    $\tuple{I, v} \gets \specialsegtree.\getperm()$\;
    \lIf{$I = \NULL$} {\Break}
    \Foreach{$o \in \optree.\mathtt{search}(I, v)$}
    {
        $\optree.\mathtt{remove}(o)$\;
        \If{\NOT $\optree.\mathtt{contains}(\valOf{o})$}
        {
            $\specialsegtree.\rmsub(\valOf{o})$\;
            $\potbotval.\push(\valOf{o})$\;
        }
    }
}
% \;
\lIf{$\potbotval = \varnothing$} { \Return $\bot$ }
\Return $\potbotval.\pop()$
}
\end{algorithm}

\subsection{Linearizability monitoring for unambiguous $\stackDS$ histories}
\seclabel{stack-optimized}

In this section, we describe our $O(n \log n)$ time algorithm
for linearizability monitoring of $\stackDS$ histories,
significantly improving upon the vanilla $O(n^4)$ time bound above,
or the algorithms in prior works~\cite{Gibbons2002LinJournal,Abdulla2025}.
Our optimized algorithm maintains a global shared
 state across multiple calls of the procedure 
\GetLinP (presented in \algoref{stack-redundant-optimized}) 
that returns a potentially bottom value if one exists,
and uses simple data structures to amortize the overall time complexity.

\myparagraph{Overview}
At a high-level, in addition to values, the algorithm also treats
individual operations, as well as \emph{partitions} of the history,
as first class concepts.
A partition $I$ of a history is simply a contiguous time interval
induced by two consecutive time points  (either invocation or response times) in the history;
there are $2n$ time points and thus $2n+1$ partitions in a history of $n$ operations.
Intuitively, each partition $I$ is an equivalence class, in the sense that,
if the span interval $(\invTimeOf{o}, \resTimeOf{o})$ of an operation $o$
intersects $I$, then in fact the entirety of $I$ is in this span, i.e., 
$I \subseteq (\invTimeOf{o}, \resTimeOf{o})$.
This allows the design of an algorithm that can work with entire
partitions at once and is a key ingredient for efficiency.
% \ucomment{
%     The algorithm iteratively determines
%     which operation $o$ is safe to remove,
%     in that, removing it (without necessarily removing all operations
%     of the value $\valOf{o}$) does not affect the linearizability status of the history (see \propref{safety-downward-closure} as the formal statement).
%     To determine this, the algorithm attempts to find
%     a partition $I$ that can witness such a safe removal of $o$.
%     At any point, if all the operations of a value $v$
%     are either completely removed in previous iterations, or are eligible to removed
%     in the most recent iteration, then $v$ is also guaranteed to be a potentially bottom
%     value.
% }
% \hcomment{
The algorithm is iterative.
In each iteration, it determines
which operation $o$ can be marked safe,
in that, it has already fulfilled the obligations of 
\defref{potentially-bottom-stack}.
To determine this, the algorithm attempts to find
a partition $I$ that can witness the safety of $o$.
We observe that a safe operation remains safe in future iterations 
(see \propref{safety-downward-closure} as the formal statement).
At any iteration, if all the operations of a value $v$
are either marked safe in previous iterations, or in the most recent iteration, 
then $v$ is also guaranteed to be a potentially bottom value.
% }
If no such value can be identified, then $\bot$ can be regarded
as a linearizability-preserving value, i,e., the stack history is non-linearizable.
The algorithm makes use of segment tree and interval tree data structures in tandem,
to store partitions and operations, to efficiently query
plausible witness partitions together with the operations corresponding to them,
and to efficiently remove such operations.
In the following we formalize when an interval is safe for an operation,
noting some helpful observations that follow.

\myparagraph{Permissive and safe partitions}
Given a stack history $\hist$, a contiguous 
interval $I = (t_a, t_b)$, and value $v\in\vals_\hist$, 
we say that $I$ is $v$-\emph{permissive} in $\hist$ if,
for every $v' \in \vals_\hist \setminus \set{v}$,
either  $\max\setpred{\invTimeOf{o}}{o\in \hist_{v'}} \leq t_a$ or 
$t_b\leq \min\setpred{\resTimeOf{o}}{o\in \hist_{v'}}$.
% \ucomment{Say history is standardized. Then simplify this and use inv($o^{push}_v$) etc. Also dont forget to remind the reader that $o^{push}_v$ means the unique push operation of v etc }
Likewise, $I$ is said to be \emph{globally permissive} in $\hist$
if it is $v$-permissive in $\hist$, for each $v \in \vals_\hist$.
Additionally, we say that $I$ is safe for an operation $o$
if $I$ is $\valOf{o}$-permissive, and further, $I \subseteq (\invTimeOf{o}, \resTimeOf{o})$.

\begin{proposition}
\proplabel{safety-downward-closure}
Let $\hist$ be a standardized unambiguous $\stackDS{}$ history,
let $o \in H$ be an operation
and let $I = (t_a, t_b)$ be a contiguous time interval.
If $I$ is safe for $o$ in $\hist$, then $I$ is also safe for $o$ in $\hist\setminus \proj{H}{V}$,
for every subset $V \subseteq \vals_\hist \setminus \set{\valOf{o}}$.
% \ucomment{
%     Further, $\hist$ is linearizable iff $\hist \setminus \set{o}$ is.
% }
\end{proposition}

\myparagraph{Data structures for efficient operations}
The set $\optree$ can be maintained as an interval tree,
and stores every operation $o$ as, indexed by its interval $(\invTimeOf{o}, \resTimeOf{o})$.
The interval tree $\optree$ supports three queries:
(a) $\optree.\mathtt{remove}(o)$ that removes operation $o$ 
and runs in time $O(\log |\optree|)$,
(b) the query $\optree.\mathtt{search}(I, v)$,
for a time interval $I$ and an (optional)
value $v \in \vals_\hist \uplus \set{\varepsilon}$,
returns
the set $S = \setpred{o \in \optree}{I \subseteq [\invTimeOf{o}, \resTimeOf{o}], v \in \set{\valOf{o}, \varepsilon}}$ in time $O(|S| \log |\optree|)$, and finally the query
(c) $\optree.\mathtt{contains}(v)$ that checks if there is any
operation of $v$ in $\optree$, and returns in time $O(1)$ time.
The set $\specialsegtree$ stores partitions and can be implemented as a 
\emph{specialized segment tree}.
It supports the following operations.
First, the query $\specialsegtree.\getperm()$ returns, 
in order of priority:
\begin{enumerate*}[label=(a)]
        \item a pair $\tuple{I, \varepsilon}$, 
        where $I$ is a globally permissive partition for $\hist$, provided one exists, or otherwise,
        \item a pair $\tuple{I, v}$, if $I$ is a $v$-permissive partition for $\hist$ , provided one exists, or
        \item $\tuple{\NULL, \varepsilon}$ if no such tuple exists.
    \end{enumerate*}
No tuple except $\tuple{\NULL, \varepsilon}$ can be returned twice.
The second query supported is
$\specialsegtree.\rmsub(v)$ that 
takes in value $v$ and updates the underlying history $H$ to $H\setminus H_v$.
In \appref{specialized-segtree} (see \algoref{stack-segtree}), we provide an efficient
implementation of this data structure.
With this implementation, for a stack history of size $n$, 
the initialization of $\specialsegtree$ takes $O(n \log{n})$ time, 
$\getperm$ takes $O(\log{n})$ time each, and $\rmsub$ takes $O(\log{n})$ time each.

Finally, we observe from \algoref{stack-redundant-optimized} that
across all the $O(|\vals_\hist|)$ calls to \GetLinP$_\stackDS$ 
(in the \algoref{decrease-conquer} framework), the total time spent is
$O(n \log n)$, where $n$ is the number of operations in the given history.
This is because, each operation and partition is touched and queried
constantly many times overall, giving us the following:

\begin{theorem}
\thmlabel{mls-dist}
    Linearizability monitoring for unamb. $\stackDS$ histories takes $O(n\log{n})$ time. \circledsmall{C1}
\end{theorem}

We implemented \algoref{stack-redundant-optimized} in our tool \fastlin, and
empirically evaluated its performance, and compare it with prior works.
See \secref{stack-empirical} for details of our evaluation.
%!TEX root=../main.tex

\section{The $\queueDS{}$ data structure}
\seclabel{queue}

\begin{figure}[t]
\specBox{
    \!\!\underline{\bf Queue data structure}: \quad $\methodsQueue = \set{\enq, \deq, \peek}$, $S_\queueDS = \vals^*$, $s_0 = \epsilon$ \\
    \vspace{-0.1in}
    \noindent
   \begin{align*}
   \begin{array}{ccccccc}
   \inferrule{s' = v \cdot s}{s \ltstrans{\opr{\enq}{v}} s'}
   &
   \;
   &
   \inferrule{\exists s'', s = s' = s'' \cdot v}{s \ltstrans{\opr{\peek}{v}} s'}
   &
   \;
   &
   \inferrule{s = s' \cdot v}{s \ltstrans{\opr{\deq}{v}} s'}
   &
   \;
   &
   \inferrule{s = s' = \epsilon}{s \ltstrans{\opr{\emp}{\empval}} s'}
   \end{array}
   \end{align*}
}
\caption{Sequential specification $\QueueSpec{}$ as an LTS.
\figlabel{queue-lts}}
\end{figure}

The sequential specification for the $\queueDS$ 
data type is presented in \figref{queue-lts}
as an LTS. 
The state of the LTS essentially encodes a queue: it $\enq$ueues in the front
but $\peek$'s and $\deq$s from the back.
Transitions capturing failed operations such as
$\deq$ or $\peek$ when $s = \epsilon$ can be modeled as $\opr{\emp}{\empval}$.
We note that its add and remove operations are: 
$\methods{}_{\queueDS}^{\mathsf{Add}} = \set{\enq}$ and
$\methods{}_{\queueDS}^{\mathsf{Remove}} = \set{\deq}$.

%!TEX root=../main.tex

\subsection{Linearizability-preserving values for unambiguous \queueDS histories}
\seclabel{queue-linearizability-preserving}

Our algorithm for monitoring \queueDS histories is similar in spirit
to \stackDS (and implicitly also fits in our decrease-and-conquer framework).
At each step, it identifies values that are \emph{potentially the first to be enqueued}
in the concurrent \queueDS history.
Consider the concurrent $\queueDS$ history on the right.
\wrapbox{r}{
  \begin{tikzpicture}[scale=0.4]
  \drawoper{2}{4}{1}{$\enq(\val{1})$}
  \drawoper{5}{7}{1}{$\peek(\val{1})$}
  \drawoper{6}{8}{2}{$\deq(\val{1})$}
  \drawoper{1}{3}{2}{$\enq(\val{2})$}
  \drawoper{9}{11}{1}{$\peek(\val{2})$}
  \drawoper{10}{12}{2}{$\deq(\val{2})$}
  \end{tikzpicture}
}{\figlabel{queue-potfront}}
Here, even though operations $\enq(\val{1})$ and $\enq(\val{2})$ are concurrent,
each of $\set{\peek(\val{1}), \deq(\val{1})}$ can possibly
be linearized before both of $\set{\peek(\val{2}), \deq(\val{2})}$, indicating
that $\val{1}$ is eligible to be enqueued before $\val{2}$, and in fact,
to be the very first value to be enqueued.
In other words, value \val{1} qualifies as a 
\emph{potentially front value}. 
In contrast, value \val{2} is certainly not a potentially front value,
as its $\peek$ operation must necessarily be scheduled after value \val{1} 
has been dequeued.

In the following, we formalize this notion in
\defref{potentially-front-queue} and justify its utility in
\lemref{linearizability-preserving-queue}.
Here, we use $o^v_{\enq}$ to denote the unique
operation on value $v$ with $\methodOf{o^v_{\enq}} = \enq$.
\begin{definition}[Potentially Front Values]
\deflabel{potentially-front-queue}
Let $\hist$ be an unambiguous $\queueDS$ history. 
A value $v \in \vals_\hist$ is said to be a potentially front value for $\hist$
if for all $v' \in \vals_\hist \setminus \set{v}$:
\begin{enumerate*}
    \item\label{queue-pfv-enq} $\invTimeOf{o^v_{\enq}} < \resTimeOf{o^{v'}_{\enq}}$, and
    \item\label{queue-pfv-others} for all $o \in \proj{\hist}{v} \cap \proj{\hist}{\set{\peek, \deq}}$ and $o' \in \proj{\hist}{v'} \cap \proj{\hist}{\set{\peek, \deq}}$, $\invTimeOf{o} < \resTimeOf{o'}$.
\end{enumerate*}
\end{definition}

\begin{restatable}{lemma}{QueueLinearizabilityPreservingLemma}
\lemlabel{linearizability-preserving-queue}
Let $\hist$ be a standardized unambiguous $\queueDS$ history and let $v \in \vals_\hist$. 
If $v$ is a potentially front value for $\hist$, then $v$ is linearizability-preserving for $\hist$.
If no such value exists, then $\bot$ is linearizability-preserving for $\hist$.
\end{restatable}
%!TEX root=../main.tex

\subsection{Linearizability monitoring for unambiguous $\queueDS$ histories}
\seclabel{queue-optimized}

% We can give an equivalent version of the definition as follows:
% \ucomment{remove}
% \begin{definition}\deflabel{potfrontval}
%   Let $H$ be a queue history. $v$ is a \emph{potentially front value} of $H$ if:
%   \begin{enumerate}
%       \item \label{queue-pfv-enq-prime} $\invTimeOf{o_{\enq(v)}} < \min\setpred{\resTimeOf{o}}{o \in \proj{H}{\enq}}$
%       \item \label{queue-pfv-others-prime} $\max\setpred{\invTimeOf{o}}{o \in \proj{H_v}{\set{\peek, \deq}}} < \min\setpred{\resTimeOf{o}}{o \in \proj{(H \setminus H_v)}{\set{\peek, \deq}}}$
%   \end{enumerate}
% \end{definition}

%!TEX root=../main.tex

\begin{algorithm}[t]
\caption{Retrieving Redundant Values for Queue Histories (Optimized)}
\algolabel{queue-redundant-optimized}
\myproc{$\init()$}{
    $S_{\propone} \gets \varnothing$ \;
    $S_{\proptwo} \gets \varnothing$ \;
}
\BlankLine
\myproc{$\GetLinP_\queueDS(H)$}{
\While{$\true$} {
	$v_1 \gets \GPFVE{H}$\;
    \If{$v_1 \neq \NULL$}{
        $S_{\propone} \gets S_{\propone} \cup \set{v_1}$\;
    	\lIf{$v_1 \in S_{\proptwo}$} {
            \Return $v_1$
        }
    }
    % \;
    $v_2 \gets \GPFVD{H}$\;
    \If{$v_2 \neq \NULL$}{
        $S_{\proptwo} \gets S_{\proptwo} \cup \set{v_2}$\;
    	\lIf{$v_2 \in S_{\propone}$} {
            \Return $v_2$
        }
    }
    % \;
	\lIf{$v_1 = v_2 = \NULL$} {\Return $\bot$}
}
}
\end{algorithm}

Based on \defref{potentially-front-queue}, a simple $O(n^2)$ implementation of
$\GetLinP$ is straightforward (see \algoref{queue-linearizability-preserving-unopt}
in \appref{queue-naive}). In turn, this gives a straightforward $O(n^3)$
algorithm for linearizability monitoring for \queueDS histories.
In this section, we instead show that this can be done in overall $O(n \log n)$ time.

\myparagraph{Overview}
Let us use the notation $\propone(\hist, v)$ to denote the
first condition in \defref{potentially-front-queue}, i.e., 
$\forall v' \neq v, \invTimeOf{o^v_{\enq}} < \resTimeOf{o^{v'}_{\enq}}$.
Similarly, let us denote by $\proptwo(\hist, v)$
to denote the second condition in \defref{potentially-front-queue}, i.e.,
for all $v' \neq v$, for all $o \in \proj{\hist}{v} \cap \proj{\hist}{\set{\peek, \deq}}$ 
and $o' \in \proj{\hist}{v'} \cap \proj{\hist}{\set{\peek, \deq}}$, $\invTimeOf{o} < \resTimeOf{o'}$.
Observe that one can independently check the truth of
$\propone(\hist, v)$ and $\proptwo(\hist, v)$.
Next observe that if $\propone(\hist, v)$ (respectively, if $\proptwo(\hist, v)$)  holds,
then $\propone(\hist \setminus \proj{\hist}{v'}, v)$ 
(resp. $\proptwo(\hist \setminus \proj{\hist}{v'}, v)$) also holds,
for every $v \neq v'$.
This in turn means that, we can track which values $v$ 
have met one, both or neither of these two properties
so far in a history $\hist$.
When a value is identified to satisfy both, it can
be deemed to be potentially front value of the history and returned.
When no such value can be found, $\bot$ is a linearizability-preserving value.

\algoref{queue-redundant-optimized} 
puts together these ideas in  a simple procedure that performs the above-mentioned checks.
Here, we maintain a set $S_{\propone}$ containing values $v$ for which
$\propone(\hist, v)$ holds up till the current residual history $\hist$.
Likewise, we maintain a set $S_{\proptwo}$ containing values $v$ for which
$\proptwo(\hist, v)$ holds up till the current residual history $\hist$.
Each invocation of $\GetLinP(H)$ accurately updates $S_{\propone}$ and 
$S_{\proptwo}$ and also determines
whether a value meets both obligations; if so, such a value is returned.
% Crucially, a value that has already been 
% added to either $S_{\propone}$ or $S_{\proptwo}$ does not need to be reconsidered.

%  obligation (but not $P_2$) for $H$, and similarly, $S_2$ for values that have satisfied $P_2$ (but not $P_1$).  
% Upon each invocation of $\GetLinP(H)$, the sets $S_1$ and $S_2$ are updated, and we check whether a value has been found that satisfies both obligations.  
% If such a value exists, it is returned as a potentially front value.  
% Additionally, any value that has already been added to either $S_1$ or $S_2$ does not need to be reconsidered.

\myparagraph{Amortizing the cost of checking \defref{potentially-front-queue}}
Instead of explicitly checking which values
satisfy either of the above properties, we make use of data structures
that internally scan timelines and track values which have an ongoing
enqueue (or ongoing peek and dequeue operations).
Our algorithm interfaces with these data structures using the procedures
$\GPFVE{H}$ and $\GPFVD{H}$ that return the next value in 
$\hist$ that satisfy resp. $\propone(\cdot, \cdot)$ and $\proptwo(\cdot, \cdot)$,
or return $\NULL$ if no remaining value does. 
No value is returned
twice by $\GPFVE{H}$ and $\GPFVD{H}$ 
across calls of $\GetLinP{H}$.
The loop (lines 3-12 in \algoref{queue-redundant-optimized}) 
runs as long as one of $v_1$ and $v_2$ is not $\NULL$, 
giving a maximum of $2|\vals_H|$ times overall across separate
calls of $\GetLinP(H)$. $\GPFVE{H}$ and $\GPFVD{H}$
can each be implemented in overall linear time, excluding initial sorting of invocation and response events of $H$.

\wrapbox{r}{
  \begin{tikzpicture}[scale=0.5]
  \drawoper{3}{5}{3}{$\enq(1)$}
  \drawhalfoper{7}{8}{3}{$\enq(4)$}
  \drawoper{1}{6}{2}{$\enq(2)$}
  \drawoper{2}{8}{1}{$\enq(3)$}
  \drawtime{4}{0.5}{5}{$t$}
  \end{tikzpicture}
}{\caption{Values $1, 2, 3$ satisfies $\propone(\cdot, \cdot)$}\figlabel{queue-minimal-enq}}

In \appref{queue-proof} (see \algoref{queue-redundant-enq}),
we outlines a possible implementation of $\GPFVE{H}$. 
The function returns values with pending 
$\enq$ operations up to the first response event of $\proj{H}{\enq}$. 
\figref{queue-minimal-enq} shows an example of time $t$ where $\NULL$
is returned until the value $1$ is removed from the history. 
It is clear that $\GPFVE{H}$ returns exactly those values satisfying $\propone(\cdot, \cdot)$ in $H$.

Ignoring the initialization and sorting of $\events$ that takes $O(n\log{n})$ time, the function takes amortized $O(1)$ time in each call and is called $O(|\vals_H|)$ times overall.

\algoref{queue-redundant-front} in \appref{queue-proof} outlines a possible implementation of $\GPFVD{H}$. The function returns values $v$ with maximum invocation event ($\resTimeOf{o^v_{\deq}}$ in a well-matched tuned history) preceding the minimum response event of $\peek$ or $\deq$ operations of other values, and returns $\NULL$ if no remaining value does. Note that the set of values in $\maxinvevents$ is maintained to be a subset of $\minresevents$'s.

Again, ignoring the initialization and sorting of $\minresevents$ and $\maxinvevents$ that takes $O(n\log{n})$ time, the function takes an amortized $O(1)$ time in each call and is called $O(|\vals_H|)$ times overall.

\begin{theorem}
  Linearizability monitoring for unamb. \queueDS histories takes $O(n \log{n})$ time. \circledsmall{C1}
\end{theorem}

In \secref{queue-empirical}, we discuss the empirical performance of \algoref{queue-redundant-optimized}
in our tool \fastlin, and how its performance compares with algorithms from prior works.
%!TEX root=../main.tex

\section{The $\pqueueDS{}$ data structure}
\seclabel{priority-queue}

\begin{figure}[t]
\specBox{
    \!\!\underline{\bf Pr. queue data structure}: $\methodsPQueue = \set{\enq, \deq, \peek}$, $S_\pqueueDS = \vals^*$, $s_0 = \epsilon$ \\
    \vspace{-0.1in}
    \noindent
   \begin{align*}
   \begin{array}{ccccccc}
   \inferrule{s' = \mathsf{sort}_{\leq_{\vals}}(v \cdot s)}{s \ltstrans{\opr{\enq}{v}} s'}
   &
   \;
   &
   \inferrule{\exists s'', s = s' = s'' \cdot v}{s \ltstrans{\opr{\peek}{v}} s'}
   &
   \;
   &
   \inferrule{s = s' \cdot v}{s \ltstrans{\opr{\deq}{v}} s'}
   &
   \;
   &
   \inferrule{s = s' = \epsilon}{s \ltstrans{\opr{\emp}{\empval}} s'}
   \end{array}
   \end{align*}
}
\caption{Sequential specification $\PQueueSpec{}$ as an LTS. $\leq_{\vals}$ is the underlying total order on values.
\figlabel{priority-queue-lts}}
\end{figure}

The sequential specification for the $\pqueueDS$ data 
type is presented in \figref{priority-queue-lts} as an LTS. 
The LTS state is a list that stores
values in increasing order according
to a fixed total order $\leq_{\vals}$ on values.
Each $\enq(v)$ operation puts the value $v$ in the appropriate place to maintain the sorted order,
whereas $\peek$ and $\deq$ operations work at the end of the list.
Transitions capturing failed operations such as
$\deq$ or $\peek$ when $s = \epsilon$ can be modeled as $\opr{\emp}{\empval}$.
We note that its add and remove operations are: 
$\methods{}_{\pqueueDS}^{\mathsf{Add}} = \set{\enq}$ and
$\methods{}_{\pqueueDS}^{\mathsf{Remove}} = \set{\deq}$.

%!TEX root=../main.tex

\subsection{Linearizability-preserving values and monitoring for unambiguous \pqueueDS histories}
\seclabel{priority-queue-linearizability-preserving}

Our monitoring algorithm for priority queues
attempts to find  linearization points
for the \peek and \deq operations of the
value with lowest priority (according to the total order $\leq_\vals$),
by checking if no other higher priority value obstructs
these operations.

% inspects the value with lowest priority (according to the total order $\leq_\vals$)
% and checks if 
% As the name suggests, for a value to be potentially minimum in a given history, it must first be of the lowest priority. Intuitively, the minimum value within a priority queue must wait for the complete removal of all higher-priority values before it can be accessed or removed.

\wrapbox{r}{
  \begin{tikzpicture}[scale=0.4]
  \drawoper{3}{4}{1}{$\enq(\val{1})$}
  \drawoper{8}{11}{1}{$\deq(\val{1})$}

  \drawoper{1}{2}{2}{$\enq(\val{2})$}
  \drawoper{5}{6}{2}{$\deq(\val{2})$}
  
  \drawoper{7}{10}{2}{$\enq(\val{3})$}
  \drawoper{12}{13}{2}{$\deq(\val{3})$}
  
  \drawtimeline{0.5}{4}{9}
  \end{tikzpicture}
}{\figlabel{pq-potlow}}

Consider the concurrent \pqueueDS{} history on the right.
Here, we assume that the priority order is $\val{1} \leq_\vals \val{2} \leq_\vals \val{3}$,
i.e., the value \val{1} has the lowest priority among all values.  
Unlike in the case of \queueDS semantics, 
the precise time at which \val{1} is being enqueued is not important,
indeed, $\val{1}$ can be enqueued at any pointm no matter what other
(higher priority) values are already present in the $\pqueueDS{}$. 
However, care must be taken when dealing with 
other operations ($\deq$ or $\peek$) of
value \val{1};
such operations can only be scheduled when no higher-priority 
values remain in the priority queue. 
Since this true of the value $\val{1}$ in the figure on the right,
we say that \val{1} is actually a \emph{potential minimum value}. 
If on the other hand, the $\deq$ operation of \val{1} was invoked 
strictly after the response of $\enq(\val{3})$, then
\val{1} would not qualify as a potential minimum value.
We formalize this definition in \defref{potentially-minimum-priority-queue}
and show that such values are, as desired, linearizability-preserving
(\lemref{linearizability-preserving-priority-queue}).

\begin{definition}[Potential Minimum Value]
\deflabel{potentially-minimum-priority-queue}
Let $\hist$ be an unambiguous $\pqueueDS$ history. 
A value $v \in \vals_\hist$ is said to be a potentially minimum value for $\hist$
if for all $o \in \proj{\hist}{v} \cap \proj{\hist}{\set{\peek, \deq}}$, 
there exists $t \in (\invTimeOf{o}, \resTimeOf{o})$ 
such that for all $v' \in \vals_\hist \setminus \set{v}$, we have
$v \leq_{\vals} v'$, and additionally one of the following holds:
\begin{enumerate*}
    \item $\max\setpred{\invTimeOf{o'}}{o' \in \proj{\hist}{v'}} < t$, or
    \item $t < \min\setpred{\resTimeOf{o'}}{o' \in \proj{\hist}{v'}}$.
\end{enumerate*}
\end{definition}

\begin{restatable}{lemma}{PriorityQueueLinearizabilityPreservingLemma}
\lemlabel{linearizability-preserving-priority-queue}
Let $\hist$ be a standardized unambiguous $\pqueueDS$ history and let $v \in \vals_\hist$. 
If $v$ is a potentially minimum value, then it is linearizability-preserving.
If no such value exists, then $\bot$ is linearizability-preserving for $\hist$.
\end{restatable}

%!TEX root=../main.tex

\begin{algorithm}[t]
\caption{Linearizability monitoring for unambiguous $\pqueueDS{}$ histories}
\algolabel{priority-queue-optimized}
\myproc{\PQLin{$\hist$}}{
$\segtree \gets \varnothing$\;
\Foreach(\tcp*[h]{in decreasing order of priority}) {$v \in \vals_\hist$}
{
    \Foreach{$o \in \proj{\hist}{v} \cap \proj{\hist}{\peek, \deq}$}
    {
        \lIf{$\segtree.\queryMin([\invTimeOf{o}, \resTimeOf{o}]) > 0$}{\Return $\false$}
    }
    \If{$\min\setpred{\resTimeOf{o}}{o\in \proj{\hist}{v'}} < \max\setpred{\invTimeOf{o}}{o\in \proj{\hist}{v'}}$}{
    $\segtree.\updateRange([\min\setpred{\resTimeOf{o}}{o\in \proj{\hist}{v'}}, \max\setpred{\invTimeOf{o}}{o\in \proj{\hist}{v'}}]$)
    }
}

\Return true
}
\end{algorithm}

We now observe that \defref{potentially-minimum-priority-queue} and 
\lemref{linearizability-preserving-priority-queue} can be rephrased as
in \corref{pqueue-final-final}.
It says that, one can simultaneously determine, for all values $v$,
if it will be eligible to be the potentially minimum value
of the history $\proj{\hist}{V}$, where $V = \setpred{v' \in \vals_\hist}{v \leq_\vals v'}$.
% Another observation here is the direct implication of \defref{potentially-minimum-priority-queue}
% that there is atmost one potentially minimum value in an unambiguous $\pqueueDS$ history $\hist$.
%  Intuitively, we can only iteratively remove the lowest priority value from $\hist$. However, we can conceptually think of ``removing'' as simply ignoring the operations of the ``removed'' values in future iterations. In fact, through the this observation, the algorithm is immediately made parallelizable, and we can check the redundance of every value simultaneously.

\begin{restatable}{corollary}{pqueuefinalfinal}\corlabel{pqueue-final-final}
  Let $H$ be a standardized unambiguous \pqueueDS{} history. 
  $H$ is linearizable iff for all $v$, and for all operations
  $o \in \proj{\hist}{v}\cap \proj{\hist}{\set{\peek, \deq}}$, 
  there exists a time point $t \in \rats_{\geq 0}$
  with $\invTimeOf{o} < t < \resTimeOf{o}$ 
  such that for all $v' \in \vals_H \setminus \set{v}$, $v \leq_\vals v'$:
  \begin{enumerate}
      \item $\max\setpred{\invTimeOf{o'}}{o' \in H_{v'}} < t$, or
      \item $t < \min\setpred{\resTimeOf{o'}}{o' \in H_{v'}}$.
  \end{enumerate}
\end{restatable}

\corref{pqueue-final-final} allows us to derive a much simpler 
procedure to monitor the linearizability of a \pqueueDS history 
compared to, say, the case of a stack, as shown in \algoref{priority-queue-optimized}.
$\segtree$ is a set of intervals supporting the following operations:
(1) $\queryMin(S)$ takes in a contiguous, but discrete segment $S$, and returns a point in $S$ which overlaps with least number of intervals
(2) $\updateRange(I)$ adds $I$ to the set.
We remark that using segment tree as an underlying data structure for $\segtree$ enables both operations to be performed in $O(\log{|\segtree|})$ time,
giving \algoref{priority-queue-optimized} its optimal $O(n \log{n})$ time complexity.

\begin{theorem}
  Linearizability monitoring for unamb. \pqueueDS histories takes $O(n \log{n})$ time. \circledsmall{C1}
\end{theorem}
%!TEX root=./main.tex

\section{Empirical Evaluation}
\seclabel{experiments}

The primary goal of our evaluation is to demonstrate the effectiveness of our log-linear
algorithms for unambiguous \setDS (\secref{set}), \stackDS (\secref{stack}),
\queueDS (\secref{queue}) and \pqueueDS (\secref{priority-queue}) data structures,
on histories derived from publicly available implementations of such data structures.
Towards this, we will evaluate 
({\bf RQ1}) whether their improved asymptotic complexity 
translates to better empirical performance compared to existing state-of-the-art tools, and 
({\bf RQ2}) whether the algorithms maintain consistent log-linear scaling behavior 
across increasingly large problem instances, establishing their suitability for 
production-scale linearizability monitoring.

\myparagraph{Implementation}
We have implemented all our algorithms in our tool \fastlin~\cite{fastlintool}.
\fastlin is primarily written in C++ to leverage its high performance and efficiency.
The input to \fastlin is a history specified as a sequence of invocation and response events,
where each event includes the details of the process that performs the event
as well as the method and the value of the corresponding operation.
We use the Scal framework~\cite{Scal2015} for obtaining implementations of concurrent data structures
as well as examples of client code that uses them.
Scal ships with suite of common implementations of data structures, including \queueDS and \stackDS. 
Scal's suite of implementations, however, does not include any implementation 
for concurrent \setDS and \pqueueDS, and for these data structures, we
implemented a custom lock-based implementation 
(as a result, the histories they generate are guaranteed to be linearizable)
and a simple client based on Scal's producer-consumer routine framework.

\myparagraph{Compared tools and setup}
We benchmark \fastlin{}'s performance against other existing 
state-of-the-art tools whose implementations are also publicly available:
Violin~\cite{Emmi2015}, Verilin~\cite{Jia2023} and the tool LiMo
due to more recent concurrent work~\cite{Abdulla2025}.
% All our benchmarks shows \fastlin performing irrefutably better in efficiency.
We remark that Violin and LiMo provides only implementations 
of linearizability checkers for \stackDS and \queueDS data types.
Several implementations of concurrent 
\stackDS and \queueDS omit $\peek$ operations,
and Violin and LiMo do not account for these operations.
We find that in absence of $\peek$ operations, 
further optimizations can significantly improve \fastlin{}'s performance, 
and we include them in our evaluation against Violin and LiMo.
For the second part of our empirical evaluation where we gauge how far \fastlin{} scales, 
we generated highly concurrent histories (20 consumers and 20 producers) 
with up to 1,000,000 operations and 
stress tested \fastlin{}'s performance against them.
We also compare its average performance (across 10 histories per data point) 
with and without optimizations for excluding $\peek$ operations. 
Our results show that the constant factor improvement 
is indeed worth considering for practical applications.
Our experiments are conducted on WSL Ubuntu 20.04 with 2.5GHz CPU and 8GB RAM.

%!TEX root=../main.tex

\subsection{Linearizability monitoring for \setDS}
\seclabel{set-empirical}

We evaluate our \setDS linearizability algorithm 
against both research questions using histories generated 
from a lock-based set implementation, which is guaranteed to generate linearizable histories.
For \textbf{RQ1} (comparative performance), 
we benchmarked \fastlin against Verilin using histories of up to 100 operations, 
with 10 histories per data point generated using a (5 producers, 5 consumers)
setup of Scal~\cite{Scal2015}. 
A timeout of $100$ seconds was set for each execution, 
and we recorded average and maximum execution times. 
Our results are summarized in~\tabref{fastlin-verilin-set}.
While Verilin frequently timed out (marked `\timeout') for histories beyond 300 operations, 
\fastlin consistently completed monitoring in under a millisecond 
across all test sizes, demonstrating substantial performance improvement.

For \textbf{RQ2} (scaling behavior validation), 
we stress-tested \fastlin on much larger histories to 
verify the log-linear complexity claims. 
\figref{fastlin-performance-set} demonstrates that
\fastlin maintains consistent scaling behavior up to one million operations, 
completing monitoring in approximately 100 milliseconds. 
The empirical scaling closely follows the theoretical log-linear bound.

\begin{figure}[t]
\centering
\begin{minipage}{0.45\textwidth}
\centering
\setlength{\tabcolsep}{3pt}
\renewcommand{\arraystretch}{0.95}
\centering
\captionof{table}{Linearizability monitoring for \setDS. \protect\circledsmall{C3}
 % --- ``\timeout'' indicates 100s timeout 
 }
\tablabel{fastlin-verilin-set}
\scalebox{0.92}{
\begin{tabular}{r | rr | rr}
\toprule
& \multicolumn{2}{c}{Verilin} & \multicolumn{2}{c}{\fastlin{}} \\
\hline
size & ave. (s)         & max (s)        & ave. (ms) & max (ms) \\ \hline
          100  & 0.066            & 0.08           & 0.0066    & 0.007    \\
          200  & 0.079            & 0.15           & 0.0086    & 0.01     \\
          300  & 0.074            & 0.09           & 0.0103    & 0.012    \\
          400  & 10.27 (\timeout) & 100 (\timeout) & 0.0107    & 0.012    \\
          500  & 40.06 (\timeout) & 100 (\timeout) & 0.012     & 0.013    \\
          600  & 60.21 (\timeout) & 100 (\timeout) & 0.0147    & 0.016    \\
          700  & 70.22 (\timeout) & 100 (\timeout) & 0.0157    & 0.018    \\
          800  & 40.23 (\timeout) & 100 (\timeout) & 0.0187    & 0.037    \\
          900  & 46.07 (\timeout) & 100 (\timeout) & 0.0286    & 0.122    \\
          1000 & 50.12 (\timeout) & 100 (\timeout) & 0.0267    & 0.036    \\
\bottomrule
\end{tabular}
}
\end{minipage}
\hfill
\begin{minipage}{0.45\textwidth}
      \scalebox{0.9}{
          \begin{tikzpicture}
          \begin{axis}[
            width=\textwidth,
            xlabel={No. of Operations},
            ylabel={Time (s)},
            xmin=0, xmax=1000000,
            ymin=0, ymax=0.15,
            legend pos=north west,
            ymajorgrids=true,
            grid style=dashed,
          ]
          \addplot table[y=set-lock(incl)] {experiments/fastlin-performance.dat};
          \end{axis}
          \end{tikzpicture}
      }
      \caption{\fastlin{} performance against \setDS histories \protect\circledsmall{C3}}
  \label{fig:fastlin-performance-set}
\end{minipage}
\vspace{0.1cm}
% \caption{Evaluating linearizability monitoring algorithms for \setDS}
\end{figure}

%!TEX root=../main.tex

\subsection{Linearizability monitoring for \stackDS}
\seclabel{stack-empirical}

\begin{table}[h]
  % \caption{Treiber stack benchmarks --- ``\timeout'' indicates 100s timeout \protect\circledsmall{C3}
    \caption{Linearizability monitoring for (Trieber) \stackDS. \protect\circledsmall{C3}
  \tablabel{fastlin-violin-verilin-treiber}
  }
    \begin{center}
    \resizebox{\textwidth}{!}{
    \scalebox{0.75}{
    \begin{tabular}{r | r r | r r | r r | r r } 
    \toprule
    & \multicolumn{2}{c}{Violin} & \multicolumn{2}{c}{Verilin} & \multicolumn{2}{c}{LiMo} & \multicolumn{2}{c}{\fastlin{}} \\ \hline
    size & ave. (s)       & max (s)        & ave. (s)       & max (s)        & ave. (s) & max (s) & ave. (s) & max (s) \\ \hline
    10k  & 100 (\timeout) & 100 (\timeout) & 100 (\timeout) & 100 (\timeout) & 8.429 & 14.17 & 0.00926 & 0.01015 \\
    20k  & 100 (\timeout) & 100 (\timeout) & 100 (\timeout) & 100 (\timeout) & 36.14 & 75.9 & 0.01977 & 0.02521 \\
    30k  & 100 (\timeout) & 100 (\timeout) & 100 (\timeout) & 100 (\timeout) & 66.60 (\timeout) & 100 (\timeout) & 0.02971 & 0.03064 \\
    40k  & 100 (\timeout) & 100 (\timeout) & 100 (\timeout) & 100 (\timeout) & 90.99 (\timeout) & 100 (\timeout) & 0.03912 & 0.04010 \\
    50k  & 100 (\timeout) & 100 (\timeout) & 100 (\timeout) & 100 (\timeout) & 100 (\timeout) & 100 (\timeout) & 0.04746 & 0.05234 \\
    60k  & 100 (\timeout) & 100 (\timeout) & 100 (\timeout) & 100 (\timeout) & 100 (\timeout) & 100 (\timeout) & 0.05980 & 0.06604 \\
    70k  & 100 (\timeout) & 100 (\timeout) & 100 (\timeout) & 100 (\timeout) & 100 (\timeout) & 100 (\timeout) & 0.06852 & 0.07301 \\
    80k  & 100 (\timeout) & 100 (\timeout) & 100 (\timeout) & 100 (\timeout) & 100 (\timeout) & 100 (\timeout) & 0.07816 & 0.08438 \\
    90k  & 100 (\timeout) & 100 (\timeout) & 100 (\timeout) & 100 (\timeout) & 100 (\timeout) & 100 (\timeout) & 0.08669 & 0.09013 \\
    100k & 100 (\timeout) & 100 (\timeout) & 100 (\timeout) & 100 (\timeout) & 100 (\timeout) & 100 (\timeout) & 0.09681 & 0.1047 \\
    \bottomrule
    \end{tabular}
    }
    }
    \end{center}
  \end{table}

We evaluate our \stackDS linearizability algorithm against
both research questions using histories generated from Scal's implementation
of the Treiber stack~\cite{thomas1986systems}.
For \textbf{RQ1} (comparative performance), 
we benchmarked \fastlin 
against three state-of-the-art tools: Violin~\cite{Emmi2015}, 
Verilin~\cite{Jia2023}, and LiMo~\cite{Abdulla2025}
using histories ranging from $10k$ to $100k$ operations; these
histories do not contain \peek operations, since none of Verilin, LiMo and Violin
implement handlers for such operations. 
We used a highly concurrent setup with multiple producers and consumers, 
setting a timeout of $100$ seconds for each execution and 
recording average and maximum execution times. 
Our results are summarized in~\tabref{fastlin-violin-verilin-treiber}. 
Violin and Verilin implement exponential time algorithms for
\stackDS linearizability, and consistently timed out across all history sizes, 
demonstrating their inability to handle moderately large concurrent histories. 
LiMo performed significantly better than them, 
successfully completing verification for smaller histories 
($10k$-$20k$ operations) but timing out beyond $30k$ operations.
We remark that LiMo implements a quadratic time algorithm for 
\stackDS linearizability monitoring.
In stark contrast, \fastlin implements our log linear algorithm and
consistently completed monitoring 
across all test sizes in under $0.1$ seconds, 
demonstrating orders-of-magnitude performance 
improvements and establishing it as the new state-of-the-art 
for linearizability monitoring for stacks.

\begin{wrapfigure}{r}{0.48\textwidth}
  \vspace{-0.2in}
    \scalebox{0.9}{
      \begin{tikzpicture}
        \begin{axis}[
        width=0.5\textwidth,
        xlabel={No. of Operations},
        ylabel={Time (s)},
        xmin=0, xmax=1000000,
        ymin=0, ymax=1.5,
        legend pos=north west,
        ymajorgrids=true,
        grid style=dashed,
        ]
        \addplot table[y=treiber(excl.)] {experiments/fastlin-performance.dat};
        \addplot table[y=kstack(excl.)] {experiments/fastlin-performance.dat};
        \addplot table[y=treiber(incl.)] {experiments/fastlin-performance.dat};
        \addplot table[y=kstack(incl.)] {experiments/fastlin-performance.dat};
      \legend{treiber (excl.), kstack (excl.), treiber (incl.), kstack (incl.)}
      \end{axis}
      \end{tikzpicture}
      }
    \caption{\fastlin{} performance against \stackDS histories \protect\circledsmall{C3}}
    \figlabel{fastlin-performance-stack}
  % \vspace{-0.2in}
\end{wrapfigure}
For \textbf{RQ2} (scaling behavior validation), 
we stress-tested \fastlin using both Treiber stack and k-stack 
implementations (available in Scal~\cite{Scal2015}) 
to generate linearizable and non-linearizable histories respectively. 
\figref{fastlin-performance-stack} demonstrates that \fastlin 
maintains consistent log-linear scaling behavior up to one million operations, 
completing monitoring in approximately one second. 
We also evaluated the impact of optimizations when 
peek operations are excluded (marked `excl.'), 
showing meaningful constant-factor improvements. 
The empirical scaling validates our theoretical complexity bounds 
and confirms the algorithm's suitability for production-scale stack verification.

%!TEX root=../main.tex

\subsection{Linearizability monitoring for \queueDS}
\seclabel{queue-empirical}

\begin{table}[h]
  \caption{Benchmark results for MS queue --- ``\timeout'' indicates a 100s timeout. \protect\circledsmall{C3}
  \tablabel{fastlin-violin-verilin-ms}}
  % \vspace{-1em} % Reduce excess vertical space
  \begin{center}
  \resizebox{\textwidth}{!}{
  \scalebox{0.75}{
  \begin{tabular}{r | r r | r r | r r | r r} 
  \toprule
  & \multicolumn{2}{c}{Violin} & \multicolumn{2}{c}{Verilin} & \multicolumn{2}{c}{LiMo} & \multicolumn{2}{c}{\fastlin{}} \\ \hline
    size  & ave. (s)       & max (s)        & ave. (s)       & max (s)        & ave. (s) & max (s) & ave. (s) & max (s) \\ \hline
    100k  & 100 (\timeout) & 100 (\timeout) & 100 (\timeout) & 100 (\timeout) & 0.484    & 0.55    & 0.05332  & 0.06847 \\
    200k  & 100 (\timeout) & 100 (\timeout) & 100 (\timeout) & 100 (\timeout) & 0.958    & 1.05    & 0.1270   & 0.1425  \\
    300k  & 100 (\timeout) & 100 (\timeout) & 100 (\timeout) & 100 (\timeout) & 1.475    & 1.63    & 0.1960   & 0.22736 \\
    400k  & 100 (\timeout) & 100 (\timeout) & 100 (\timeout) & 100 (\timeout) & 2.107    & 2.27    & 0.2537   & 0.3097  \\
    500k  & 100 (\timeout) & 100 (\timeout) & 100 (\timeout) & 100 (\timeout) & 2.669    & 2.85    & 0.3505   & 0.3912  \\
    600k  & 100 (\timeout) & 100 (\timeout) & 100 (\timeout) & 100 (\timeout) & 3.087    & 3.3     & 0.3840   & 0.4058  \\
    700k  & 100 (\timeout) & 100 (\timeout) & 100 (\timeout) & 100 (\timeout) & 3.613    & 3.76    & 0.4435   & 0.5151  \\
    800k  & 100 (\timeout) & 100 (\timeout) & 100 (\timeout) & 100 (\timeout) & 4.005    & 4.16    & 0.5406   & 0.5972  \\
    900k  & 100 (\timeout) & 100 (\timeout) & 100 (\timeout) & 100 (\timeout) & 4.067    & 4.44    & 0.6966   & 0.7971  \\
    1M & 100 (\timeout) & 100 (\timeout) & 100 (\timeout) & 100 (\timeout) & 5.404    & 5.56    & 0.7333   & 0.7910  \\
    \bottomrule
    \end{tabular}
    }
  }
  \end{center}
  % \vspace{-1em} % Reduce excess vertical space
\end{table}

\begin{wrapfigure}{r}{0.5\textwidth}
  \scalebox{0.9}{
    \begin{tikzpicture}
    \begin{axis}[
      width=0.5\textwidth,
      xlabel={No. of Operations},
      ylabel={Time (s)},
      xmin=0, xmax=1000000,
      ymin=0, ymax=0.75,
      legend pos=north west,
      ymajorgrids=true,
      grid style=dashed,
    ]
    \addplot table[y=ms(excl)] {experiments/fastlin-performance.dat};
    \addplot table[y=bs-kfifo(excl)] {experiments/fastlin-performance.dat};
    \addplot table[y=ms(incl.)] {experiments/fastlin-performance.dat};
    \addplot table[y=bs-kfifo(incl)] {experiments/fastlin-performance.dat};
    \legend{ms (excl), bs-kfifo (excl), ms (incl.), bs-kfifo (incl)}
    \end{axis}
    \end{tikzpicture}
  }
  \caption{\fastlin{} performance against \queueDS histories \protect\circledsmall{C3}}
  \figlabel{fastlin-performance-queue}
  \vspace{-0.3in}
\end{wrapfigure}
We evaluate our \queueDS linearizability algorithm 
following the same methodology as \secref{stack-empirical}, 
using histories generated from Scal's implementation 
of the Michael-Scott queue~\cite{Michael1996}. 
For \textbf{RQ1}, 
we benchmarked \fastlin against Violin, Verilin, and LiMo 
using histories ranging from $100k$ to $1M$ operations (larger than in stack,
given that at least LiMo implements an $O(n \log n)$ time algorithm).
As summarized in~\tabref{fastlin-violin-verilin-ms}, Violin and Verilin 
again timed out across all sizes, 
while LiMo showed notably better performance on queues 
than stacks, successfully completing all test cases but still requiring several 
seconds for larger histories (5.4 seconds for $1M$ operations). 
\fastlin maintained its superior performance, completing $1M$ operations in only 0.73 seconds.

For \textbf{RQ2}, we used Michael-Scott queue and bounded-size k-FIFO 
queue implementations to generate linearizable 
and non-linearizable histories respectively. 
\figref{fastlin-performance-queue} confirms consistent log-linear 
scaling up to one million operations, 
with similar constant-factor improvements when peek operations are excluded.

%!TEX root=../main.tex

\subsection{Linearizability monitoring for \pqueueDS}
\seclabel{priority-queue-empirical}

We evaluate our \pqueueDS linearizability algorithm 
using the same experimental setup as~\secref{set-empirical}, 
with histories generated from a lock-based priority queue implementation in Scal. 
For \textbf{RQ1}, we benchmarked \fastlin against Verilin on histories up to $1000$ operations
(with timeout of 100 seconds). 
As shown in~\tabref{fastlin-verilin-priority-queue}, 
Verilin timed out for histories beyond $100$ operations, 
while \fastlin consistently completed monitoring in under 2 milliseconds. 
For \textbf{RQ2}, \figref{fastlin-performance-priority-queue}
demonstrates consistent log-linear scaling up to one million operations, 
completing in approximately one second.

%!TEX root=../main.tex

\begin{figure}[h]
\centering
\begin{minipage}{0.45\textwidth}
\centering
\setlength{\tabcolsep}{3pt}
\renewcommand{\arraystretch}{0.95}
\centering
\captionof{table}{Linearizability monitoring for \pqueueDS. \protect\circledsmall{C3}
 % --- ``\timeout'' indicates 100s timeout 
 }
\tablabel{fastlin-verilin-priority-queue}
\scalebox{0.85}{
\begin{tabular}{r | rr | rr}
\toprule
& \multicolumn{2}{c}{Verilin} & \multicolumn{2}{c}{\fastlin{}} \\
\hline
size & ave. (s)          & max (s)        & ave. (ms) & max (ms) \\ \hline
          100  & 0.184             & 0.89           & 0.1193    & 0.130    \\
          200  & 41.175 (\timeout) & 100 (\timeout) & 0.2269    & 0.251    \\
          300  & 60.053 (\timeout) & 100 (\timeout) & 0.2787    & 0.352    \\
          400  & 80.029 (\timeout) & 100 (\timeout) & 0.3346    & 0.358    \\
          500  & 80.03 (\timeout)  & 100 (\timeout) & 0.4039    & 0.454    \\
          600  & 60.064 (\timeout) & 100 (\timeout) & 0.4355    & 0.508    \\
          700  & 90.015 (\timeout) & 100 (\timeout) & 0.5760    & 0.629    \\
          800  & 90.016 (\timeout) & 100 (\timeout) & 0.6165    & 0.687    \\
          900  & 90.017 (\timeout) & 100 (\timeout) & 0.6827    & 0.758    \\
          1000 & 90.024 (\timeout) & 100 (\timeout) & 1.1596    & 1.687    \\
\bottomrule
\end{tabular}
}
\end{minipage}
\hfill
\begin{minipage}{0.5\textwidth}
      \scalebox{0.9}{
          \begin{tikzpicture}
          \begin{axis}[
            width=\textwidth,
            xlabel={No. of Operations},
            ylabel={Time (s)},
            xmin=0, xmax=1000000,
            ymin=0, ymax=1.0,
            legend pos=north west,
            ymajorgrids=true,
            grid style=dashed,
          ]
          \addplot table[y=pq-lock(excl)] {experiments/fastlin-performance.dat};
          \addplot table[y=pq-lock(incl)] {experiments/fastlin-performance.dat};
          \legend{pq-lock (excl), pq-lock (incl)}
          \end{axis}
          \end{tikzpicture}
      }
      \caption{\fastlin{} performance against histories of \pqueueDS \protect\circledsmall{C3}}
  \figlabel{fastlin-performance-priority-queue}
\end{minipage}
\vspace{0.1cm}
% \caption{Evaluating linearizability monitoring algorithms for \pqueueDS}
\end{figure}

%!TEX root=./main.tex

\section{Related Work and Discussion}

\begin{comment}
 - Complexity of monitoring
 	1. How hard is race prediction
 	2. Complexity of weak memory testing
 	3. Complexity of isolation level testing
 	4. Against regular languages (Sabadini, Mathur)
\end{comment}

\myparagraph{Linearizability and its monitoring}
Gibbons and Korach~\cite{Gibbons1997}
initiated the complexity-theoretic study of linearizability,
showing $\np$-hardness in general and tractability for unambiguous registers.
Given the intractability, a thorough complexity-theoretic understanding
of this problem has remained elusive for a broader class of
data structures, except for the work of \cite{Emmi2018},
which, as we show in~\secref{collection-types} has incorrect results.
In~\cite{Gibbons1997,Gibbons2002LinJournal} Gibbons et al studied
the linearizability monitoring problems for stacks, queues, priority queues
and other data structures in the setting when distinct values are inserted
in the data structure, and more recently
Abdulla et al~\cite{Abdulla2025} also study the problem for stacks, queues, sets
and multisets when the input histories are \emph{data differentiated}, and also propose
efficient algorithms.
While there is overlap in some of our results and that 
of~\cite{Abdulla2025,Gibbons1999Lin,Gibbons2002LinJournal}, our work improves these results
on several fronts, as we discuss in~\secref{closely-related}. 
Recent practical tools for monitoring take the stress testing 
approach~\cite{lincheck2023,lincheck2024},
where for a given implementation of a concurrent data structure, one looks
for a large number of executions to see if any of them violate linearizability,
and for the sake of flexibility, implement generic algorithms that
enumerate exponentially many interleavings, either explicitly~\cite{Jia2023,Burckhardt2010,Wing1993,Lowe2017,Zhang2015, Horn2015} or symbolically~\cite{relinche2025, Emmi2015},
and occasionally using small depth heuristics~\cite{ozkan2019checking}.
Besides the classic notion of linearizability proposed by Herlihy and Wing~\cite{Wing1990},
other notions such as strong linearizability~\cite{StrongLinearizability2011},
quantitative relaxations of specifications~\cite{Henzinger2013} as well as 
durable linearizability~\cite{DurableLinearizability2016} have been proposed.
% to cater for
% preservation of hyperproperties, gain performance and ensure correctness under crashes.

\myparagraph{Automated and deductive verification for linearizability}
On the front of checking entire programs, decidability
results have been shown, including
for that against regular specifications~\cite{Alur2000,Hamza2015} for bounded
domain programs or restrictions on how the data structures are invoked~\cite{vcerny2010model}, 
outside of which, undecidability persists~\cite{Bouajjani2013}. 
Nevertheless, semi-decidable techniques have been extensively studied~\cite{DongolDerrickSurvey2015}.
Several deductive proof systems for verifying linearizability for data structure
algorithms and implementations~\cite{LinearizabilityForward2024,JungProphecy2019,AspectOriented2013},
but require significant manual annotation.
%and expertise. 

\myparagraph{Monitoring other concurrency-centric properties}
With recent industrial adoption of model checking,
the core monitoring questions for concurrent programs and their algorithmic
and complexity-theoretic aspects have become popular.
The work on verifying sequential consistency by Gibbons and Korach~\cite{Gibbons1997} has now inspired
further work including sub-classes~\cite{cantin2005complexity,abdulla2018optimal,Agarwal2021},
other memory models~\cite{gontmakher2003complexity,chakraborty2024hard,tuncc2023optimal,chini2020framework},
database isolation levels~\cite{biswas2019complexity,AWDITMoldrup2025}, 
distributed systems consistency levels~\cite{Bouajjani2017causal,biswas2019complexity} 
and runtime predictive 
analysis~\cite{Mathur2020,kulkarniCONCUR,mathur2021optimal,grains2024,AngPredictive2024,OSR2024Shi,Shi2025MessagePassing,Ang2025GPrefix,Mathur2020Atomicity,Kini2017WCP,Mathur2022PredictiveTutorial,Pavlogiannis2019M2,TuncDeadlock2023}.

\subsection{Detailed comparison with Gibbons et al~\cite{Gibbons1999Lin} and Abdulla et al~\cite{Abdulla2025}}
\seclabel{closely-related}

Gibbons et al~\cite{Gibbons2002LinJournal}, and more recently Abdulla et al~\cite{Abdulla2025}
also study the linearizability monitoring problem for stacks (both), queues (both), 
priority queues (\cite{Gibbons2002LinJournal}) and sets (\cite{Abdulla2025})
and propose similar tractability results under similar restrictions like unambiguity.
Apart from our generic decrease-and-coquer algorithmic framework,
there are significant differences between the our log linear algorithms and 
their algorithms for the same data structures.
The first difference comes from the choice of operations.
Notably, in our work we consider histories
with peek/contains operations as well as `failed' operations 
(which we denote using $\mempty$ operations), which is in sharp contrast to both prior
works~\cite{Gibbons2002LinJournal,Abdulla2025}.
In absence of peek operations, the unambiguity restriction boils down to
the more restricted version where each pair of method name and 
value appears exactly once (`data differentiated'~\cite{Abdulla2025}),
where there is a unique pair of `matching' operations per value.
This simplicity allows for the design of algorithms that 
take constantly many $O(n)$ passes on the input.
We think that both sets of algorithms~\cite{Gibbons2002LinJournal,Abdulla2025}
cannot be easily extended to include such operations. 
Our algorithm naturally, yet carefully, accounts for peek events, 
in overall $O(n\log{n})$ time. 
Intuitively, disallowing peek/contains operations in is analogous 
to disallowing multiple reads of the same value in a register history, 
and thus unnatural; we also remark that Gibbons and Korach do account 
for multiple reads of the same value in their very first work~\cite{Gibbons1997}. 
Likewise, exteding the algortihmic insights of~\cite{Abdulla2025,Gibbons1999Lin}
to account for `failed' operations also appears non-trivial.
The second significant improvement is due to the complexity of 
monitoring for stacks. 
The algorithm in~\cite{Abdulla2025} works in $O(n^2)$ time. 
The algorithm in~\cite{Gibbons2002LinJournal}
works in $O(n^3)$ time and is fundamentally different from ours because
it addresses a more general problem (VSC problem) when the input history does not
necessarily induce an \emph{interval order}. In this setting, when one also considers
peek operations, the monitoring problem  becomes $\np$-hard
since there is a simple polynomial time many-one
reduction from the VSC problem of unambiguous registers
to VSC problem of unambiguous stacks.

%!TEX root=./main.tex

\subsection{Collection types and OOS Linearizability~\cite{Emmi2018}}
\seclabel{collection-types}

In 2018, Emmi and Enea proposed a tractability result 
for linearizability monitoring of unambiguous histories of a class
of  data types called \emph{collection types}, defined by a semantic characterization~\cite{Emmi2018}.
This tractability argument presented is two-folds:
First, it is shown that, for collection types, each instance of the linearizability
monitoring problem for unambiguous histories can be mapped in polynomial time
to an instance of another problem ---- OOS linearizability (for a fixed quantifier rank).
The input to an OOS Linearizability problem is a first order logic formula
(`Operation Order Specification' or OOS for short), that encodes
the happens-before constraints of the given history, together with the
constraints of the sequential specification of the collection type in question,
and the linearizability problem boils down to checking if the formula
is satisfiable in a model, equipped with a total order that is consistent with 
the  happens-before constraint encoded in the formula.
Second, it is argued that
OOS linearizability can be solved in polynomial time for OOS'es of fixed quantifier ranks.
We revisit the second step and argue that OOS linearizability (even for quantifier rank of $\leq 3$)
is $\np$-hard (in contrast to~\cite{Emmi2018}), 
and preset the full proof in~\appref{collection-types}:

\begin{restatable}{theorem}{OOSLinearizabilityNPHardness}
\thmlabel{oos-linearizability-np-hard}
OOS Linearizability is $\np$-hard.
\end{restatable}

We also identify the error in the (incorrect) tractability
claim of the OOS linearizability problem, and identified 
\emph{Hornification} as the source of unsoundness.
In essence, an OOS is
equisatisfiable with its (quantifier-free) ground instantiation (which
is only polynomially longer when the quantifier rank of the OOS is fixed)
that only contains happens-before constraints.
Hornification then asks to come transform this resulting quantifier-free formula
into a Horn formula, which is then argued to result into an equisatisfiable formula.
In contrast we argue that Hornification does not preserve satisfiability
and localize the issue in the proof of the equi-satisfaction claim
(see \appref{collection-types}).
In summary, tractability results for a generic class of
data structures (such as collection types) remains an open problem.
Our proposal of container types in \secref{containers}
(which include priority queues, that are not collection types)
is a step towards obtaining a generic algorithm for linearizability monitoring, though
tractable algorithms continue to remain elusive.

%!TEX root=./main.tex

\section{Conclusions and Future Work}

In this work, we developed a unified decrease-and-conquer framework for the problem of linearizability monitoring, and instantiated it to obtain efficient algorithms for unambiguous {\registerDS}s, {\setDS}s, {\stackDS}s, {\queueDS}s, and {\pqueueDS}s. Under the unambiguity restriction, we show that these fundamental data structures admit log-linear time monitors, thereby improving on the best-known complexity bounds and offering a significant performance advantage in practice. We further validated our algorithms through an implementation in \fastlin, demonstrating scalability to million-operation histories and outperforming existing state-of-the-art tools.

Beyond these contributions, our work opens several new avenues. A natural direction is to extend the decrease-and-conquer paradigm to richer concurrent objects, such as maps, dequeues, mutexes and counters. Another interesting direction is to relax the unambiguity restriction and explore whether variants of our framework can handle ambiguous or value-reuse histories without incurring exponential cost.

\section*{Data Availability Statement}
Our tool \fastlin is publicly available at~\cite{fastlintool}, and the version used in this work is available at~\cite{zenodoArtifact}.

\begin{acks}
This work is partially supported by the National Research Foundation, Singapore, and Cyber Security Agency of Singapore under its National Cybersecurity R\&D Programme (Fuzz Testing <NRF-NCR25-Fuzz-0001>). Any opinions, findings and conclusions, or recommendations expressed in this material are those of the author(s) and do not reflect the views of National Research Foundation, Singapore, and Cyber Security Agency of Singapore. 
\end{acks}

\bibliographystyle{ACM-Reference-Format}
\bibliography{references}

% \clearpage

% \appendix

%!TEX root=./main.tex

\appendix

%!TEX root=../main.tex

\section{Common Lemmas}

\begin{lemma}[Tightening Lemma]
\lemlabel{tight-hist}
    Let $H$ be a history and $t_1 < t_2$ be two rational numbers such that for all $o \in H$, $t_1 < \resTimeOf{o}$ and $\invTimeOf{o} < t_2$. Given any linearization $\lin$, there exists another linearization $\lin'$ such that $\tau_\lin = \tau_{\lin'}$ and $t_1 < \lin'(o) < t_2$ for all $o \in H$.
\end{lemma}
\begin{proof}
    Let ${\sf res}_{\min} = \min\setpred{\resTimeOf{o}}{o \in H}$ be the minimum response time in $H$. Let $S = \set{o_1, o_2, ..., \allowbreak o_k}$ be the maximal subset of operations for which $\lin(o) \leq {\sf res}_{\min}$ for $o \in S$, and $\lin(o_1) < \lin(o_2) < ... < \lin(o_k)$. By assumption, $t_1 < {\sf res}_{\min}$. We construct an assignment of time $\lin'$ where $\lin'(o) = \lin(o)$ for $o \in H \setminus S$, and $\lin'(o_i) = \max\set{t_1 + \frac{i}{|S|+1}({\sf res}_{\min} - t_1),\lin(o_i)}$ for $o \in S$. Notice that $\invTimeOf{o} < \lin(o) \leq \lin'(o) < {\sf res}_{\min} \leq \resTimeOf{o}$ for all $o \in S$. Hence, $\lin$ is indeed a linearization of $H$. Secondly, $\tau_\lin$ and $\tau_{\lin'}$ remains the same (i.e. $\lin'(o_1) < \lin'(o_2) < ... < \lin'(o_k) < {\sf res}_{\min}$). Lastly, $t_1 < \lin'(o)< {\sf res}_{\min}$ for $o \in S$.

    Symmetrically, we can construct a reassignment $\lin''$ from $\lin'$ for operations in the set $\set{o\in H \;|\; \lin'(o) > \max\set{\invTimeOf{o} \;|\; o \in H}}$ such that $\lin''(o) < t_2$ for $o \in S$. The conclusion follows.
\end{proof}

\begin{lemma}[Flushing Lemma]\label{flush-hist}
    Let $H$ be a history and $t_1 < t_2$ be two rational numbers such that for all $o \in H$, $t_1 < \resTimeOf{o}$ and $\invTimeOf{o} < t_2$. Given any linearization $\lin$, there exists linearizations $\lin'$ and $\lin''$ such that:
    \begin{enumerate}
        \item $\tau_\lin = \tau_{\lin'} = \tau_{\lin''}$,
        \item $\lin'(o) \leq \min\set{t_2, \lin(o)}$ for all $o \in H$, and
        \item $\max\set{t_1, \lin(o)} \leq \lin''(o)$ for all $o \in H$.
    \end{enumerate}
\end{lemma}
\begin{proof}
    By \lemref{tight-hist}, there exists legal linearization $\lin'''$ such that $\tau_\lin = \tau_{\lin'''}$ and $t_1 < \lin'''(o) < t_2$ for all $o \in H$. We then construct $\lin'$ where $\lin'(o) = \min\set{\lin'''(o), \lin(o)}$ for all $o \in H$. Clearly, $\lin'(o) \leq \min\set{t_2, \lin(o)}$. Secondly, $\lin'$ must be a linearization since $\invTimeOf{o} \leq \lin'''(o), \lin(o) \leq \resTimeOf{o}$.

    Let $o_1, o_2 \in H$ be two operations such that $\lin(o_1) < \lin(o_2)$, and thus, $\lin'''(o_1) < \lin'''(o_2)$. Suppose $\lin'(o_1) \geq \lin'(o_2)$. By construction, $\lin'(o_1)$ is one of $\lin(o_1)$ and $\lin'''(o_1)$. Without loss of generality, assume $\lin'(o_1) = \lin(o_1)$. Then it must be the case that $\lin'(o_2) \leq \lin'(o_1) = \lin(o_1) < \lin(o_2)$. Since $\lin'(o_2) < \lin(o_2)$, we have $\lin'(o_2) = \lin'''(o_2)$ (again, $\lin'(o_2)$ is one of $\lin(o_2)$ and $\lin'''(o_2)$ by construction). However, since $\lin'''(o_1) < \lin'''(o_2)$, we have that $\lin'''(o_1) < \lin'(o_1) = \min\set{\lin'''(o_1), \lin(o_1)}$, contradiction. Hence, $\lin'$ preserves the order of the abstract sequence (i.e. $\lin'(o_1) < \lin'(o_2)$ for all $o_1, o_2 \in H$ where $\lin(o_1) < \lin(o_2)$).

    Symmetrically, $\lin''$ can be constructed in the same way. The conclusion follows.
\end{proof}
\clearpage
%!TEX root=../main.tex

\section{Proofs from \secref{decrease-and-conquer}}
\seclabel{app-containers}

\begin{proposition}
\proplabel{collections-are-containers}
The class of collection types~\cite{Emmi2018} is included
in the class of container types (\defref{container-type}).
\end{proposition}

\begin{proof}
Let $\spec_\ADT$ be a collection type, and $s \in \spec_\ADT$ be some operation sequence. If $\proj{s}{\vals \setminus \set{v}} = \varepsilon$, we are done as $\varepsilon \in \spec_\ADT$. Now, let $s' \cdot o = \proj{s}{\vals \setminus \set{v}}$ for some $v \in \vals$, operation sequence $s'$ and operation $o$.
Let $s'' \cdot o$ be a prefix of $s$. As specifications are prefix-closed, $s'' \cdot o \in \spec_\ADT$. Assume $\drop{s'', \set{v}} \cdot o \notin \spec_\ADT$. By definition, $\uses(s''\cdot o, \set{v})$ holds. Hence, $\uses'(o, \set{v})$ holds, where $\uses'$ is the local interpretation of $\uses$. Hence, $\touches(o, \set{v})$ holds. By locality, since operations touch exactly the values appearing in its label, $v = \valOf{o}$, contradicting the assumption that $s' \cdot o = \proj{s}{\vals \setminus \set{v}}$.
Hence, $\drop{s'', \set{v}} \cdot o = s' \cdot o \in \spec_\ADT$. The conclusion follows.
\end{proof}
\clearpage
%!TEX root=../main.tex

\section{Proofs from \secref{standardization}}
\applabel{app-standardization}

\begin{lemma}\lemlabel{remove-helper}
Let $\ADT \in \ADTReducedSet$.
Let $\hist$ be a well-matched unambiguous and invocation-response-time-compliant history of $\ADT$. 
$\hist$ is linearizable iff $\hist\setminus \hist_\empval$ is linearizable, 
and further, for each $o \in \hist_\empval$,
there exists $V \subseteq \vals_\hist \setminus \set{\empval}$,
two subhistories $\hist_{o, 1} = \proj{\hist}{V}$
and $\hist_{o, 2} = \hist \setminus \hist_\empval \setminus \hist_{o, 1}$, and a time
$t \in (\invTimeOf{o}, \resTimeOf{o})$, that splits across $\hist_{o,1}$ and $\hist_{o, 2}$, i.e.,   
\begin{enumerate*}
    \item for every $o' \in \hist_{o, 1}$, $\invTimeOf{o'} < t$, and
    \item for all $o' \in \hist_{o, 2}$, $t < \resTimeOf{o'}$.
\end{enumerate*}
\end{lemma}
\begin{proof}
    \begin{itemize}
        \item[($\Rightarrow$)] $H\setminus H_\empval$ is linearizable by 
        \propref{cotainer-linearizability-per-value}. Assume a legal linearization $\lin$. For each $o \in H\setminus H_\empval$, let $t = \lin(o)$. It is clear that $t\in(\invTimeOf{o}, \resTimeOf{o})$. Let $H_1' = \set{o'\in H \;|\; \lin(o') < \lin(o)}$ and $H_2' = H \setminus H_1'\setminus \set{o}$. $H_1'$ and $H_2'$ must intuitively be well-matched. Additionally, for each $o'\in H_1'$, $\invTimeOf{o'} < \lin(o') < t$. Finally, for each $o' \in H_2'$, $t < \lin(o') < \resTimeOf{o'}$.

        \item[($\Leftarrow$)] Let $H_\empval = \set{o_1, o_2, ..., o_k}$, and $t_i$ be a time that satisfy the above two properties for operation $o_i$, $1 \leq i \leq k$. Without loss of generality, assume $t_i < t_{i+1}$ for all $1 \leq i < k$. It is clear that for each $1 \leq i \leq k$, $H_{o_i}' = H_{o_i, 1} \setminus \bigcup_{1\leq j<i}{H_{o_j}'}$ is also well-matched.
        Since $H_{o_i}' \subseteq H \setminus H_\empval$ and $H \setminus H_\empval$ is linearizable, $H_{o_i}'$ must also be linearizable. Notice that for all $o\in H_{o_i}'$, $t_{i-1} < \resTimeOf{o}$ and $\invTimeOf{o} < t_i$. Hence, there exists a legal linearization $\lin_i$ of $H_{o_i}'$ such that $t_{i-1} < \lin_i(o) < t_i$ by \lemref{tight-hist} (assume $t_0$ to be 0). Similarly, there must also exist a legal linearization $\lin_{k+1}$ of $H_{k+1}' = H \setminus \bigcup_{1\leq j \leq k}{H_{o_j}'}$ such that $t_k < \lin_{k+1}(o)$ for all $o \in H_{k+1}'$. Let $\lin_\empval$ be a linearization of $H_\empval$ where $\lin_\empval(o_i) = t_i$. Reader may verify that the union of the linearizations $\lin_\empval, \lin_1, \lin_2, ..., \lin_{k+1}$ is indeed a legal linearization of $H$.
    \end{itemize}
\end{proof}

\removeEmptyOperations*
\begin{proof}
    Corollary follows directly from \lemref{remove-helper}.
\end{proof}
\clearpage
%!TEX root=../main.tex

\section{Proofs from \secref{set}}
\seclabel{set-proof}

\SetLinearizabilityPreservingLemma*

\begin{proof}
\begin{itemize}
  \item[($\Rightarrow$)]
  Assume legal linearization $\lin$ of $\proj{H}{v}$. It is clear that for all $o \in \proj{H}{v} \setminus \proj{H}{v}^\true$, we must have that either $\lin(o) < \lin(o_{\minsert(v)})$ or $\lin(o_{\remove(v)}) < \lin(o)$. Hence, either $\invTimeOf{o} < \lin(o) < \lin(o_{\minsert(v)}) < \resTimeOf{o_{\minsert(v)}}$ or $\resTimeOf{o} > \lin(o) > \lin(o_{\remove(v)}) > \invTimeOf{o_{\remove(v)}}$.
  
  \item[($\Leftarrow$)]
  Suppose that the given conditions hold, we construct a legal linearization $\lin$. Assume $\resTimeOf{o_{\minsert(v)}} < \invTimeOf{o_{\remove(v)}}$. We safely schedule operations in $o\in \proj{H}{v} \setminus \proj{H}{v}^\true$ either before $\resTimeOf{o_{\minsert(v)}}$ or after $\invTimeOf{o_{\remove(v)}}$, and we can do that by assumption. Thereafter, we schedule operations in $\proj{H}{v}^\true$ within the interval $(MinRes - \epsilon, MaxInv + \epsilon)$ for some sufficiently small rational number $\epsilon$, and we can do so due to $\proj{H}{v}^\true$ being a tuned subhistory and \lemref{tight-hist}.
  
  Now assume $\resTimeOf{o_{\minsert(v)}} \geq \invTimeOf{o_{\remove(v)}}$, we then safely schedule operations in $o\in \proj{H}{v} \setminus \proj{H}{v}^\true$ either before or after $\resTimeOf{o_{\minsert(v)}}$. Again,operations in $\proj{H}{v}^\true$ are concurrent and safely schedulable within $(\resTimeOf{o_{\minsert(v)}} - \epsilon, \resTimeOf{o_{\minsert(v)}} + \epsilon)$ for some sufficiently small $\epsilon$ by \lemref{tight-hist}.
\end{itemize}
\end{proof}

\clearpage
%!TEX root=../main.tex

\section{Proofs from \secref{stack}}
\applabel{stack-proof}

%!TEX root=../main.tex

\begin{algorithm}
\caption{Linearizability-preserving values for unambiguous $\stackDS$ histories (unoptimized)}
\algolabel{stack-redundant}
\myproc{$\GetLinP_\stackDS(H)$}{

\Foreach{$v \in \vals_H$}
{
    $IsPotBot = \true$\;
    \Foreach{$o \in H_v$}
    {
        $I \gets (\invTimeOf{o}, \resTimeOf{o})$\;
        \Foreach{$v' \in \vals_H \setminus \set{v}$}
        {
            \If{$\min\setpred{\resTimeOf{o'}}{o' \in H_{v'}} < \max\setpred{\invTimeOf{o'}}{o' \in H_{v'}}$}
            {
                $I \gets I \cap (\min\setpred{\resTimeOf{o'}}{o' \in H_{v'}}, \max\setpred{\invTimeOf{o'}}{o' \in H_{v'}})$\;
            }
        }
        \lIf{$I = \varnothing$}
        {
            $IsPotBot = \false$
        }
    }
    \lIf{$IsPotBot = \true$} { \Return $v$ }
}

\Return $\bot$
}
\end{algorithm}

In this section, we use $\potbotval(H)$ to denote the set of potentially bottom values in $H$.

\begin{proposition}\proplabel{stack-reduce-well-matched}
	Given a linearizable stack history $H$. Any well-matched subhistory $H' \subseteq H$ is also linearizable.
\end{proposition}

\begin{lemma}\lemlabel{stack-sound}
    Given a stack history $H$. For all $v \in \potbotval(H)$, $v$ is linearizability-preserving.
\end{lemma}
\begin{proof}
    It is clear from \defref{container-type} that if $H$ is linearizable, so will $H \setminus \proj{H}{v}$. It suffices to show that if $H \setminus \proj{H}{v}$ is linearizable, so will $H$. We prove by construction. Let $\lin'$ be a legal linearization of $H \setminus \proj{H}{v}$.
    
    Let $\proj{H}{v} = \set{o_1, o_2, ..., o_k}$ and $t_1, t_2, ..., t_k$ be times satisfying, for each respectively operation, the condition of \defref{potentially-bottom-stack}. We can assume that $\methodOf{o_1} = \push$, $\methodOf{o_k} = \pop$, and $t_i < t_{i+1}$ for all $1 \leq i < k$. We can realize the assumption with any given set of timings by swapping $t_1$ and $t_k$ with the current lowest and the current highest timings respectively if not already. We construct a linearization $\lin_v$ of $\proj{H}{v}$ with $\lin_v(o_i) = t_i$ for all $1 \leq i \leq k$. By the non-triviality of $H$ (see \secref{standardization}), $\lin_v$ is a legal linearization of $\proj{H}{v}$.
    
    Let $G_i$ denote the union of the set $\set{\proj{H}{v_{i1}}, \proj{H}{v_{i2}}, ..., \proj{H}{v_{ik_i}}}$ for which $\invTimeOf{o'} < t_i$ for all $o' \in \proj{H}{v_{ij}}$, $1 \leq i \leq k$, $1 \leq j \leq k_i$. It is clear that for each $1 \leq i \leq k$, $G_i' = G_i \setminus \bigcup_{1\leq j < i}{G_j}$ is also well-matched, and since $G_i' \subseteq H \setminus \proj{H}{v}$, it is also linearizable by \propref{stack-reduce-well-matched}. Notice that for all $o'\in G_i'$, $t_{i-1} < \resTimeOf{o'}$ and $\invTimeOf{o'} < t_i$. Hence, there exists a legal linearization $\lin_i$ of $G_i'$ such that $t_{i-1} < \lin_i(o') < t_i$ by \lemref{tight-hist} (assume $t_0$ to be 0). Similarly, there must also exist a legal linearization $\lin_{k+1}$ of $G_{k+1}' = H \setminus \bigcup_{1\leq j \leq k}{G_j}$ such that $t_k < \lin_{k+1}(o)$ for all $o \in G_{k+1}'$.
        
	Finally, let $\lin$ the union of the linearizations $\lin_v, \lin_1, \lin_2, ..., \lin_{k+1}$. Reader may verify that $\lin$ is indeed a legal linearization of $H$.
\end{proof}

\begin{lemma}\lemlabel{stack-sufficient}
    Given a non-empty linearizable stack history $H$, $\potbotval(H) \neq \varnothing$.
\end{lemma}
\begin{proof}
	Let $\lin$ be a legal linearization of $H$. We have $v$ be the first value pushed into the stack in $\tau_\lin$. It is clear that since the history is well-matched, we have that for all $o \in \proj{H}{v}$, and $v' \in \vals_H \setminus \set{v}$, one of the following holds:
    \begin{enumerate}
        \item for all $o' \in H_{v'}$, $\lin(o') < \lin(o)$, or
        \item for all $o' \in H_{v'}$, $\lin(o) < \lin(o')$.
    \end{enumerate}
	By corollary,
	\begin{enumerate}
        \item for all $o' \in H_{v'}$, $\invTimeOf{o'} < \lin(o)$, or
        \item for all $o' \in H_{v'}$, $\lin(o) < \resTimeOf{o'}$.
    \end{enumerate}
    Hence, $\max\setpred{\invTimeOf{o}}{o\in \proj{H}{v'}} < \lin(o)$ or $\lin(o) < \min\setpred{\resTimeOf{o}}{o\in \proj{H}{v'}}$. Hence, $v \in \potbotval(H)$.
\end{proof}

\stackredundant*
\begin{proof}
    The soundness of $\potbotval$ follows directly from \lemref{stack-sound}. Assuming non-empty $H$, if $\potbotval(H) = \varnothing$, $H$ is not linearizable by \lemref{stack-sufficient}. Hence $\bot$ is linearizability preserving.
\end{proof}

\subsection{Implementing $\specialsegtree$ as a specialized segment tree}
\applabel{specialized-segtree}

%!TEX root=../main.tex

\begin{algorithm}
\caption{Internal implementation of $\specialsegtree$ using a segment tree $S$}
\algolabel{stack-segtree}
\myproc{$\init(H)$}{
\Foreach{$v \in \vals_H$}
{
  \tcp{$\waitingreturns$ is implemented as a hashmap that maps values to a set of partitions}
  $\waitingreturns\gets\varnothing$ \; 
  \tcp{$\pendingreturns$ is implemented as a hashset of partitions that supports a \deq operation for removing and returning an arbitrary element}
  $\pendingreturns\gets\varnothing$ \;
  $S$.updateRange($I_v^c$, $\tuple{1, v}$)\tcp{Assume $\vals_H\subseteq \nats$}
}
}
\;
\myproc{$\rmsub(v)$}{
  $S$.updateRange($I_v^c$, $\tuple{-1, -v}$)\;
  $\pendingreturns\gets\pendingreturns\cup\setpred{I_{pos}}{\tuple{I_{pos}, v'} \in \waitingreturns, v = v'}$\;
}
\BlankLine
\myproc{$\getperm()$}{
  \If{$\pendingreturns \neq \varnothing$} {\Return $\tuple{\pendingreturns.\deq(), \varepsilon}$}
  \;
  $\tuple{pos, \tuple{layers, v}}  \gets S$.queryMin()\;
  $S$.updateRange([$pos$, $pos$], $\tuple{\infty, \infty}$)\;
  \If(\tcp*[h]{The unique partition containing $pos$ is globally permissive for the current history}) {$layers = 0$}
  {
      return $\tuple{I_{pos}, \varepsilon}$
  }
  \If(\tcp*[h]{The unique partition containing $pos$ is  $v$-permissive for the current history}) {$layers = 1$}
  {
      $\waitingreturns\gets\waitingreturns\cup\set{\tuple{I_{pos}, v}}$\;
      return $\tuple{I_{pos}, v}$
  }
  return $\tuple{\NULL, \varepsilon}$\;
}
\end{algorithm}

$\specialsegtree$ in \algoref{stack-redundant-optimized} is 
intuitively a wrapper around a base segment tree $S$
that (efficiently) stores (possibly overlapping)
weighted intervals, together a set of values associated with them, 
and internally represents these intervals as weighted discrete points (each of which is also associated with a value).
We show its internal implementation in \algoref{stack-segtree}.
The query $S$.updateRange$(I, \tuple{k, v})$ updates the set of all discrete
points in the interval $I$ by incrementing their weights with $k \in \nats$
and incrementing the original value $v'$ by $v$ (to the final value $v+v'$).
The query $S$.queryMin() returns the discrete point with weight $layers$ and value $v$
such that its weight $layers$ is the lowest amongst all discrete points.
In \algoref{stack-segtree}, we use the term
$I_v^c$ to denote the interval $[\min\setpred{\resTimeOf{o}}{\in H_{v'}}, \max\setpred{\invTimeOf{o}}{o\in H_{v'}}]$.

\clearpage
%!TEX root=../main.tex

\section{Details, omitted from \secref{queue}}
\applabel{queue-proof}

In this section, we use $\potfrontval(H)$ to denote the set of potentially front values in $H$.

\begin{lemma}\lemlabel{queue-sound}
    Given a queue history $H$. For all $v \in \potfrontval(H)$, $v$ is linearizability-preserving.
\end{lemma}
\begin{proof}
    It is clear from \defref{container-type} that if $H$ is linearizable, so will $H \setminus \proj{H}{v}$. It suffices to show that if $H \setminus \proj{H}{v}$ is linearizable, so will $H$. We prove by construction. Let $\lin'$ be a legal linearization of $H \setminus \proj{H}{v}$.

    We let $t_1$ be some rational number such that $\max\setpred{\invTimeOf{o}}{o \in \proj{\proj{H}{v}}{\set{\peek, \deq}}} < t_1 < \min\setpred{\resTimeOf{o}}{o \allowbreak \in \proj{(H \setminus \proj{H}{v})}{\set{\peek, \deq}}}$. By Lemma \ref{flush-hist}, there exists linearization $\lin'$ such that:
    \begin{enumerate}
        \item $\proj{\tau_{\lin'}}{\set{\peek, \deq}} = \proj{\tau_\lin}{\set{\peek, \deq}}$,
        \item $\proj{\tau_{\lin'}}{\enq} = \proj{\tau_\lin}{\enq}$,
        \item $\max\set{t_1, \lin(o)} \leq \lin'(o)$ for all $o\in \proj{(H\setminus \proj{H}{v})}{\set{\peek, \deq}}$, and
        \item $\lin'(o) = \lin(o)$ for all $o\in\proj{(H\setminus \proj{H}{v})}{\enq}$.
    \end{enumerate}
    Since the order of $\enq$ operations remains the same in $\lin'$ and similarly for $\peek$ and $\deq$ operations (with some $\peek$ and $\deq$ operations linearized later), $\lin'$ is still a legal linearization of $H\setminus \proj{H}{v}$.

    Now, let $t_2$ be some rational number such that $\invTimeOf{o_{\enq(v)}} < t_2 < \min\setpred{\resTimeOf{o}}{o \in \proj{H}{\enq}}$. Since $H$ is assumed to be a tuned history, $t_2 < \min\setpred{\resTimeOf{o}}{o \in H\setminus \proj{H}{v}}$. Again by Lemma \ref{flush-hist}, there exists linearization $\lin''$, $\tau_{\lin''} = \tau_{\lin'}$, such that $\max\set{t_2, \lin'(o)} \leq \lin''(o)$ for all $o\in H\setminus \proj{H}{v}$. It is clear that $\lin''$ is still a legal linearization of $H\setminus \proj{H}{v}$.
    
    Now we can safely assign $\lin''(o_{\enq(v)})$ such that $\lin''(o_{\enq(v)}) < t_2$, and similarly $\lin''(o) < t_1$ for all $o \in \proj{\proj{H}{v}}{\set{\peek, \deq}}$. Reader may verify that $\lin''$ is indeed a legal linearization of $H$ where $v$ is the first enqueued, peeked, and dequeued value. The conclusion follows.
\end{proof}

\begin{lemma}\lemlabel{queue-sufficient}
    Given a non-empty linearizable queue history $H$, $\potfrontval(H) \neq \varnothing$.
\end{lemma}
\begin{proof}
    Let $\lin$ be a legal linearization of $H$, and $v$ be the first enqueued value in $\tau_\lin$. We have that for all $v' \in \vals_H \setminus \set{v}$:
    \begin{enumerate}
        \item $\lin(o_{\enq(v)}) < \lin(o_{\enq(v')})$, and
        \item for all $o_v \in \proj{H_v}{\set{\peek, \deq}}$ and $o_{v'} \in \proj{H_{v'}}{\set{\peek, \deq}}$, $\lin(o_v) < \lin(o_{v'})$.
    \end{enumerate}
    By corollary,
    \begin{enumerate}
        \item $\invTimeOf{o_{\enq(v)}} < \resTimeOf{o_{\enq(v')}}$, and
        \item for all $o_v \in \proj{H_v}{\set{\peek, \deq}}$ and $o_{v'} \in \proj{H_{v'}}{\set{\peek, \deq}}$, $\invTimeOf{o_v} < \resTimeOf{o_{v'}}$.
    \end{enumerate}
    Hence, $v \in \potfrontval(H)$.
\end{proof}

\QueueLinearizabilityPreservingLemma*
\begin{proof}
    The soundness of $\potfrontval$ follows directly from \lemref{queue-sound}. Assuming non-empty $H$, if $\potfrontval(H) = \varnothing$, $H$ is not linearizable by \lemref{queue-sufficient}. Hence $\bot$ is linearizability preserving.
\end{proof}

\subsection{Naive polynomial time algorithm for monitoring \queueDS histories}
\applabel{queue-naive}

%!TEX root=../main.tex

\begin{algorithm}[t]
\caption{Linearizability-preserving values for unambiguous $\queueDS$ histories (unoptimized)}
\algolabel{queue-linearizability-preserving-unopt}
\myproc{$\GetLinP_\queueDS(H)$}{

\Foreach{$v \in \vals_H$}
{
    $IsPotFront \gets \true$\;
    \Foreach{$v' \in \vals_H \setminus \set{v}$}
    {
        \If{$\invTimeOf{o_{\enq(v)}} >= \resTimeOf{o_{\enq(v')}}$}
        {
            $IsPotFront \gets \false$\;
        }
    }
    \Foreach{$o \in \proj{H_v}{\set{\peek, \deq}}$}
    {
        \Foreach{$v' \in \vals_H \setminus \set{v}$}
        {
            \Foreach{$o' \in \proj{H_{v'}}{\set{\peek, \deq}}$}
            {
                \If{$\invTimeOf{o} >= \resTimeOf{o'}$}
                {
                    $IsPotFront \gets \false$\;
                }
            }
        }
    }
    \lIf{$IsPotFront = \true$} { \Return $v$ }
}

\Return $\bot$
}
\end{algorithm}

\algoref{queue-linearizability-preserving-unopt} shows the simple, unoptimized 
implementation of \GetLinP based on \defref{potentially-front-queue}. 
We return any potential front value, or $\bot$ if there are none. The correctness of the algorithm is guaranteed by \lemref{linearizability-preserving-queue}. 
It is easy to see that it runs in polynomial time.
% The tractability of \algoref{queue-redundant} brings us directly the following theorem.

% \begin{restatable}{theorem}{queuelin}\thmlabel{queue-lin}
% Linearizability monitoring for unambiguous queue histories can be solved in polynomial time.
% \end{restatable}

\subsection{Auxiliary procedures for monitoring \queueDS histories}
\applabel{queue-opt}

%!TEX root=../main.tex

\begin{algorithm}
\caption{Retrieving Values Satisfying $\propone(\cdot, \cdot)$}
\algolabel{queue-redundant-enq}
\myproc{$\init()$}{
$\events \gets \varnothing$\;
}
\BlankLine
\myproc{\GPFVE{$H$}}{
    \If(\tcp*[h]{initialization of sorted events queue}){$\events = \varnothing$} {
        \For{$o \in \proj{H}{\enq}$} {
            $\events \gets \events \cup \set{\tuple{\invTimeOf{o}, o},\tuple{\resTimeOf{o}, o}}$
        }
    }
    \;
    \While{$\events \neq \varnothing$} {
        $\tuple{t, o} \gets \events.\peek()$\;
        \lIf{$t = \resTimeOf{o}$ \AND $o\in \proj{H}{\enq}$}{\Return $\NULL$}
        $\events.\deq()$\;
        \lIf{$t = \invTimeOf{o}$}{\Return $\valOf{o}$}
    }
}
\end{algorithm}
%!TEX root=../main.tex

\begin{algorithm}
\caption{Retrieving Values Satisfying $\proptwo(\cdot, \cdot)$}
\algolabel{queue-redundant-front}
\myproc{$\init()$}{
$\minresevents \gets \varnothing$\;
$\maxinvevents \gets \varnothing$\;
}
\BlankLine
\myproc{\GPFVD{$H$}}{
    \If(\tcp*[h]{initialization of sorted events queues}){$\minresevents = \varnothing$} {
        \For{$v \in \vals_H$} {
            $\minresevents \gets \minresevents \cup \set{\tuple{\min\setpred{\resTimeOf{o}}{o \in \proj{H_v}{\set{\peek, \deq}}}, v}}$\;
            $\maxinvevents \gets \maxinvevents \cup \set{\tuple{\max\setpred{\invTimeOf{o}}{o \in \proj{H_v}{\set{\peek, \deq}}}, v}}$
        }
    }
    \;
    \tcp{only first two minimum response events used in this iteration}
    \While{$\minresevents.\first().\second() \notin \vals_H$}{
        $\minresevents.\deq()$
    }
    \While{$|\minresevents| > 1$ \AND $\minresevents.\second().\second() \notin \vals_H$}{
        $\minresevents \gets \minresevents \setminus \set{\minresevents.\second()}$
    }
    \;
    \lIf{$\maxinvevents = \varnothing$}{\Return $\NULL$}
    \;
    \If{$\maxinvevents.\first().\first() < \minresevents.\first().\first()$}{
        \Return $\maxinvevents.\deq().\second()$\;
    }
    \;
    $v \gets \minresevents.\first().\second()$\;
    $t \gets \max\setpred{\invTimeOf{o}}{o \in \proj{H_v}{\set{\peek, \deq}}}$\;
    \If{$\tuple{t, v} \in \maxinvevents$ \AND $(|\minresevents| = 1$ \OR $t < \minresevents.\second().\first())$}{
        $\maxinvevents \gets \maxinvevents \setminus \set{\tuple{t, v}}$\;
        \Return $v$
    }
    \;
    \Return $\NULL$
}
\end{algorithm}

\clearpage
%!TEX root=../main.tex

\section{Proofs for \secref{priority-queue}}\seclabel{pqueue-proof}

In this section, we use $\potlowval(H)$ to denote the set of potentially minimum values in $H$.

\begin{proposition}\proplabel{priority-queue-reduce-well-matched}
	Given a linearizable priority queue history $H$. Any well-matched subhistory $H' \subseteq H$ is also linearizable.
\end{proposition}

\begin{lemma}\lemlabel{priority-queue-sound}
    Given a priority queue history $H$. For all $v \in \potlowval(H)$, $v$ is linearizability-preserving.
\end{lemma}
\begin{proof}
    It is clear from \defref{container-type} that if $H$ is linearizable, so will $H \setminus \proj{H}{v}$. It suffices to show that if $H \setminus \proj{H}{v}$ is linearizable, so will $H$. We prove by construction. Let $\lin'$ be a legal linearization of $H \setminus \proj{H}{v}$.

    Let $\proj{H_v}{\set{\peek, \deq}} = \set{o_1, o_2, ..., o_k}$ where $\methodOf{o_k} = \deq$, and $t_1, t_2, ..., t_k$ be the time points satisfying the properties of \defref{potentially-minimum-priority-queue}.
    
    We can assume that $t_i < t_{i+1}$ for all $1 \leq i < k$. We can realize the assumption with any given set of timings by swapping $t_k$ with the current highest timing if not already. We construct a linearization $\lin_v$ of $H_v$ with $\lin_v(o_i) = t_i$ for all $1 \leq i \leq k$, and $\lin_v(o_{\minsert(v)})$ before all. By the non-triviality of $H$ (see \secref{standardization}), $\lin_v$ is a legal linearization of $H_v$.
    
    Let $G_i$ denote the union of the set $\set{\proj{H}{v_{i1}}, \proj{H}{v_{i2}}, ..., \proj{H}{v_{ik_i}}}$ for which $\invTimeOf{o'} < t_i$ for all $o' \in \proj{H}{v_{ij}}$, $1 \leq i \leq k$, $1 \leq j \leq k_i$. Without loss of generality, assume $t_i < t_{i+1}$ for all $1 \leq i < k$. It is clear that for each $1 \leq i \leq k$, $G_i' = G_i \setminus \bigcup_{1\leq j < i}{G_j}$ is also well-matched, and since $G_i' \subseteq H \setminus H_v$, it is also linearizable by \propref{priority-queue-reduce-well-matched}. Notice that for all $o'\in G_i'$, $t_{i-1} < \resTimeOf{o'}$ and $\invTimeOf{o'} < t_i$. Hence, there exists a legal linearization $\lin_i$ of $G_i'$ such that $t_{i-1} < \lin_i(o') < t_i$ by \lemref{tight-hist} (assume $t_0$ to be 0). Similarly, there must also exist a legal linearization $\lin_{k+1}$ of $G_{k+1}' = H \setminus \bigcup_{1\leq j \leq k}{G_j}$ such that $t_k < \lin_{k+1}(o)$ for all $o \in G_{k+1}'$.
        
	Finally, let $\lin$ the union of the linearizations $\lin_v, \lin_1, \lin_2, ..., \lin_{k+1}$. Reader may verify that $\lin$ is a legal linearization of $H$.
\end{proof}

\begin{lemma}\lemlabel{priority-queue-sufficient}
    Given a non-empty linearizable priority queue history $H$, $\potlowval(H) \neq \varnothing$.
\end{lemma}
\begin{proof}
	Let $\lin$ be a legal linearization of $H$. We have $v = \min_{<_\nats}{\vals_H}$ be the value of least priority. We have that for all $o \in \proj{\proj{H}{v}}{\set{\peek, \deq}}$ and $v' \in \vals_H \setminus \set{v}$, one of the following holds:
    \begin{enumerate}
        \item for all $o' \in \proj{H}{v'}$, $\lin(o') < \lin(o)$, or
        \item for all $o' \in \proj{H}{v'}$, $\lin(o) < \lin(o')$.
    \end{enumerate}
	By corollary,
	\begin{enumerate}
        \item for all $o' \in \proj{H}{v'}$, $\invTimeOf{o'} < \lin(o)$, or
        \item for all $o' \in \proj{H}{v'}$, $\lin(o) < \resTimeOf{o'}$.
    \end{enumerate}
    Hence, $\max\setpred{\invTimeOf{o}}{o\in \proj{H}{v'}} < \lin(o)$ or $\lin(o) < \min\setpred{\resTimeOf{o}}{o\in \proj{H}{v'}}$. Hence, $v \in \potlowval(H)$.
\end{proof}

\PriorityQueueLinearizabilityPreservingLemma*
\begin{proof}
    The soundness of $\potlowval$ follows directly from \lemref{priority-queue-sound}. Assuming non-empty $H$, if $\potlowval(H) = \varnothing$, $H$ is not linearizable by \lemref{priority-queue-sufficient}. Hence $\bot$ is linearizability preserving.
\end{proof}

\clearpage
%!TEX root=../main.tex

\section{Recap on OOS Linearizability for Collections Types}
\applabel{collection-types}

In 2018, Emmi and Enea~\cite{Emmi2018} proposed a tractability result in the
context of linearizability monitoring of unambiguous histories of a class
of  data types called \emph{collection types}.
This tractability argument proceeds by showing that,
(1) for collection types, each instance of the linearizability
monitoring problem for unambiguous histories can be mapped in polynomial time
to an instance of another problem ---- OOS (`Operation Order Specification')
linearizability (for a fixed quantifier rank), and
(2) OOS linearizability can be solved in polynomial time for OOS'es of fixed quantifier ranks.
In this section, we present a brief overview of OOS linearizability
and show that this problem is in fact $\np$-hard.
We next precisely identify the error in the (now known to be incorrect) 
proof of tractability of this problem~\cite[Theorem 5.10]{Emmi2018}\footnote{We communicated the details of the problem with this prior result to the authors of~\cite{Emmi2018} more than a year ago, via email. 
The authors have acknowledged the reciept of the email, but have not confirmed the problem with the proof yet.}.

\subsection{OOS Linearizability and its Intractability}

At a high level, an OOS specification $\varphi$ is a
many-sorted first order logic formula
over sorts $\opSort$, $\methodSort$, $\valSort$.
The vocabulary of such formulae comes from
a \emph{history signature}, that includes of 
(1) a dedicated binary relation symbol $\befrel : \opSort \times \opSort$,
(2) a dedicated function symbol $\mthd: \opSort \to \methodSort$, and
(3) a dedicated function symbol $\vl: \opSort \to \valSort$, and
(4) appropriate constant symbols, such as symbols corresponding to method names.
A model (or structure) of such an OOS formula $\varphi$ 
is a tuple $\mdl = (U_{\opSort}, U_{\methodSort}, U_{\valSort}, \interp)$
that consists of three disjoint universes, one for each sort,
as well as an interpretation $\interp$ that maps relation and function symbols
to relations and functions of appropriate types, and such that
$\interp(\befrel) \subseteq U_{\opSort} \times U_{\opSort}$ is a partial order.
It is straightforward how a history $\hist$ can be translated to a model $\mdl_{\hist}$.
$\mdl$ is said to be a \emph{total} model if $\interp(\befrel)$ is a total order.
A model $\mdl' = (U'_{\opSort}, U'_{\methodSort}, U'_{\valSort}, \interp')$ 
extends model $\mdl = (U_{\opSort}, U_{\methodSort}, U_{\valSort}, \interp)$ if 
they only (possibly) differ in the interpretation of $\befrel$, which additionally satisfies
$\interp(\befrel) \subseteq \interp'(\befrel)$.
A model $\mdl$ is said to be \emph{safe} for an OOS formula $\varphi$ 
if there is an extension $\mdl'$ of it that is total and further satisfies $\mdl' \models \varphi$.
A history $\hist$ is said to be \emph{OOS linearizable} for an OOS formula $\varphi$
if $\mdl_{\hist}$ is safe for $\varphi$.
Emmi and Enea show that, for unambiguous histories of a collection type $\ADT$,
one can derive an OOS formula (of fixed quantifier rank)
$\varphi_{\ADT}$ so that $\hist$ is OOS linearizabile for $\varphi_{\ADT}$  
iff $\hist$ is linearizable.
Here, in contrast to~\cite{Emmi2018}, we show that the OOS linearizability is in fact
an $\np$-hard problem.

\OOSLinearizabilityNPHardness*

\begin{proof}
    We show that there is a polynomial time many-one reduction from the
    problem of linearizability monitoring of register histories (known to be $\np$-hard\cite{Gibbons1997}) 
    to the OOS linearizability problem.
    For each instance $\hist$ of the register linearizability problem,
    we construct an instance $(\hist', \varphi)$ in polynomial time
    so that $\hist'$ is a register history and further
    $\hist$ is linearizable iff $\hist'$ is OOS linearizable with respect to $\varphi$. 
    We simply choose $\hist' = \hist$.
    The formula $\varphi$ is chosen to be the following formula, that intuitively captures
    the sequential specification of a 
    register\footnote{While we only present here a simplified formulation of an OOS specification, the
    original formulation asks for a more elaborate input $\tuple{\Sigma, I, \varphi, R}$ which we present here for the sake of completeness: our signature $\Sigma$ is clear from 
    the formula, the interpretation is the \emph{standard interpretation}~\cite{Emmi2018},
   $\varphi$ is as described above, and $R$ is the identity binary relation over histories}:
    \begin{align*}    
        \varphi\!\equiv\!\forall x, (\mthd(x) = \mread)\!\Rightarrow\!\exists y,
        \bigg(
        \begin{array}{c}\mthd(y) = \mwrite \land \vl(x)=\vl(y) \land \bef{y, x} \\ 
        \land
            \nexists z, \big(\mthd(z) = \mwrite \land \bef{y, z} \land \bef{z, x}\big)
            \end{array}\bigg)
    \end{align*}
    It is easy to see that the reduction takes linear time.
    Let us now argue that the reduction is correct. Consider a history $\hist$.
    If $\hist$ is linearizable, then there is a linearization $\lin$ such that
    the corresponding abstract history satisfies $\tau_\lin \in \RegisterSpec$.
    It is easy to see that the extended history $\hist_\lin$ of corresponding to
    $\lin$ satisfies $\varphi$ since in it, every read operation is preceded by 
    a write operation of the same value, without being intervened by another write operation.
    This means $\mdl_\hist$ is safe for $\varphi$.
    In the other direction, assume that $\mdl_\hist$ is safe for $\varphi$.
    This means that there is a model $\mdl'_\hist$ that satisfies $\varphi$, is total and extends
    $\mdl_\hist$. Observe that any total model can naturally be interpreted as a
    linearization, and we denote by $\lin$, the linearization corresponding to $\mdl'_\hist$.
    Since $\mdl'_\hist$ extends $\mdl_\hist$, we must also have that $\lin$ is a linearization of $\hist$.
    Finally, since $\mdl'_\hist \models \varphi$, we must have that
    $\lin$ must be a member of $\RegisterSpec$. 
    This means that $\hist$ is a  linearizable register history.
\end{proof}

\subsection{Isolating the error in the proof: Hornification is unsound}

Here, we investigate deeper and identify where exactly the proof of tractability
of OOS linearizability in \cite{Emmi2018} breaks.
The crux of the argument in \cite{Emmi2018} is two folds.
First, given an OOS formula $\varphi$ and a history $\hist$, one can derive
a ground (i.e., quantifier-free) formula 
in CNF $\psi_\hist = \varphi_\hist \land \varphi_{\textsf{total}}$
containing only the relation symbol $\befrel$;
here $\varphi_\hist$ is obtained by instantiating the quantifiers from
$\varphi$ using the model $\mdl_\hist$ of the history $\hist$,
while the formula $\varphi_{\textsf{total}}$ denotes that $\befrel$ is a total
order and is obtained by instantiating the total order axioms using $\mdl_\hist$.
More importantly, it is shown that $\hist$ is OOS linearizable for $\varphi$
iff $\psi_\hist$ is satisfiable (in the pure first order sense)~\cite[Lemma 5.5]{Emmi2018}.
Following this, it is shown that the resulting ground CNF formula 
$\psi_\hist$ can be translated
to an equisatisfiable ground Horn formula $\psi_\hist^{\textsf{Horn}}$
using a \emph{Hornification scheme}
but we think this claim (see~\cite[Lemma 5.9]{Emmi2018}) is incorrect --- Hornification
does not preserve satisfiability as we show below.

\myparagraph{Hornification does not preserve satisfiability}
Let us first recall the notion of Hornification scheme due to~\cite{Emmi2018}.
Recall that the first step of ground instantiation
of an OOS formula results into a ground CNF formula whose literals are of either the form
$l = \bef{o_a, o_b}$ or $l = \neg \bef{o_a, o_b}$, 
where $o_a$ and $o_b$ represent constant symbols, 
intuitively representing operations from the history in consideration.
Let us define $\sim$ to be the smallest equivalence on such literals so that
for every $o, o'$, we have $\bef{o, o'} \sim \neg \bef{o', o}$.
This equivalence can then be lifted to equal-length clauses 
$C = (l_1 \lor l_2 \ldots l_m)$ and $C' = (l'_1 \lor l_2 \ldots l'_m)$
as follows:
$C \sim C'$ iff $\bigwedge l_i \sim l'_i$.
The Hornification of a ground clause $C$ is then the formula
 $C^{\mathsf{Horn}} = \bigwedge\limits_{C' \sim C,\; C' \text{ is a Horn clause}} C'$,
 and the Hornification of a ground formula $\psi = C_1 \land C_2 \ldots C_k$ is
the formula $\psi^{\mathsf{Horn}} = C^{\mathsf{Horn}}_1 \land C^{\mathsf{Horn}}_2 \ldots C^{\mathsf{Horn}}_k$.

In essence, Hornification removes the non-Horn clauses
(which are imminent due to totality axioms) and replace them with all
$\sim$-equivalent Horn clauses, resulting into a Horn formula.
However, such a translation does not preserve satisfiability.
Consider the formula $\psi = C_1 \land C_2 \land C_3 \land C_4$
over two constants $o_1$ and $o_2$, where
\begin{align*}
\begin{array}{lcl}
C_1 = (\neg \bef{o_1, o_2} \lor \bef{o_2, o_1}) & \quad & C_2 =  (\bef{o_1, o_2} \lor \neg \bef{o_2, o_1}) \\ 
C_3 = (\neg \bef{o_1, o_2} \lor \neg \bef{o_2, o_1}) & \quad & C_4 = (\bef{o_1, o_2} \lor \bef{o_2, o_1})
\end{array}
\end{align*}
Observe that the last two clauses $C_3$ and $C_4$ together encode totality.
The Hornification of $\psi$ will contain exactly the clauses $C_1, C_2$ and $C_3$,
which are satisfiable simultaneously (by setting both $\bef{o_1, o_2}$ as well as $\bef{o_2, o_1}$ to false)!
This contradicts~\cite[Lemma 5.9]{Emmi2018}.

\myparagraph{Localizing the error}
We localized the error in the proof of \cite[Lemma 5.9]{Emmi2018} to the 
`\textbf{Base case}' which claims that
for any resolution proof `$\psi \entails_0 C$' comprising of $0$ non-Horn resolvents,
one can derive a resolution proof `$\psi^{\mathsf{Horn}} \entails C'$' for some $C' \sim C$.
More precisely, the following statement~\cite[beginning of page 16]{Emmi2018} is incorrect:
``\emph{Since $c_2$ cannot contain other positive literal than $b$ and $b \sim \neg a$, 
the conclusion $b \lor c_2$ of this rule is included in $F$.}'' as it overlooks the case
when the clause $(c_2 \lor \neg a)$ may be obtained by previous applications of the resolution rule,
in which case it may not be present in the formula $\bar{F}$ 
obtained by hornification of $F$.    

\end{document}